\documentclass[preprint,12pt]{elsarticle}

\pdfoutput=1

\usepackage{amssymb}

\journal{Carbon 153 (2019) 242-256}

\begin{document}

\begin{frontmatter}

\title{Yielding and jerky plasticity of tilt grain boundaries in high-temperature graphene \tnoteref{t1}}
\tnotetext[t1]{This manuscript version is made available under the CC BY-NC-ND license
  https://creativecommons.org/licenses/by-nc-nd/4.0/}

\author[mymainaddress]{Wenquan Zhou}
\author[mymainaddress]{Jincheng Wang\corref{mycorrespondingauthor}}
\cortext[mycorrespondingauthor]{Corresponding author}
\ead{jchwang@nwpu.edu.cn}
\author[mymainaddress]{Bo Lin}
\author[mymainaddress]{Zhijun Wang}
\author[mymainaddress]{Junjie Li}
\author[mysecondaryaddress]{Zhi-Feng Huang\corref{mycorrespondingauthor}}
\ead{huang@wayne.edu}
\address[mymainaddress]{State Key Laboratory of Solidification Processing,
  Northwestern Polytechnical University, Xi'an 710072, China}
\address[mysecondaryaddress]{Department of Physics and Astronomy, Wayne State University,
  Detroit, Michigan 48201, USA}

\begin{abstract}
Graphene is well known for its extraordinary mechanical properties combining
brittleness and ductility. While most mechanical studies of graphene focused on
the strength and brittle fracture behavior, its ductility, plastic deformation,
and the possible brittle-to-ductile transition, which are important for
high-temperature mechanical performance and applications, still remain much
less understood. Here the mechanical response and deformation dynamics of
graphene grain boundaries are investigated through a phase field crystal
modeling, showing the pivotal effects of temperature and local dislocation
structure. Our results indicate that even at relatively high temperature
(around $3350$ K), the system is still governed by a brittle fracture and
cracking dynamics as found in previous low-temperature experimental and
atomistic studies. We also identify another type of failure dynamics with
low-angle grain boundary disintegration. When temperature increases a transition
to plastic deformation is predicted. The appearance of plastic flow at ultrahigh
temperature, particularly the phenomenon of jerky plasticity, is attributed to
the stick and climb-glide motion of dislocations around the grain boundary.
The corresponding mechanism is intrinsic to two-dimensional systems, and governed
by the competition between the driving force of accumulated local stress and the
defect pinning effect, without the traditional pathways of dislocation generation
needed in three-dimensional materials.
\end{abstract}

\end{frontmatter}

\section{Introduction}

Graphene has attracted tremendous amounts of interest in terms of both fundamental research
and applications. One of the advantages of its related device applications is the exceptional
mechanical properties of this two-dimensional (2D) material, such as ultrahigh strength and
elastic modulus and also high flexibility \cite{Geim1530,Lee385}. However, these properties
are affected by the fact that, during the growth and fabrication of large scale graphene films
that are required in most applications, topological defects, including dislocations and grain
boundaries (GBs), are inevitably introduced. Recent experimental studies demonstrated that
the intrinsic strength and fracture behavior of graphene monolayers are influenced by these
defects particularly GBs \cite{HuangNature11,LeeScience13,doi:10.1021/nl200429f,KimNanoLett12,
RasoolNatCommun13}, such as the reduction of fracture strength and the direction of crack
propagation. A significant amount of effort has been devoted to the study of GB structures
and their effects on mechanical response of the system, as driven by the requirements of
material mechanical performance in graphene-based devices and also by the need for
understanding and predicting the fundamental mechanisms of defects and for achieving the
sample property control.

Most mechanical experiments were conducted around room temperature, showing brittleness
of sample failure in graphene polycrystals. On the other hand, graphene has been known
to incorporate both brittleness and ductility \cite{Geim1530}, and its ultrahigh melting
temperature (around $5000$ K \cite{SinghPRB13,LosPRB15}) is particularly attractive for
potential high-temperature applications. Although graphene sheets, especially those with
defects, are susceptible to oxidation at high temperature, in many device applications
graphene is integrated or encapsulated inside as a functional layer without being exposed
to ambient conditions \cite{MayorovNanoLett11}, which is important for further extension
across a wide range of temperature. In addition, any generic mechanisms identified
for high-temperature 2D systems would be useful for the study of a broader variety of
novel 2D material applications such as graphene oxide paper \cite{DikinNature07} and
various graphene-based functionalized hybrid materials \cite{Geim1530,SestSEMSC18,
SaniSEMSC18,AbinayaSEMSC18}. Some applications of graphene-type materials at high
temperature have already been explored. One example is the 2D light-emitting device
made of graphene \cite{KimNatNanotechnol15}, where a temperature as high as $2800$ K
has been imposed to emit visible light from graphene, while mechanical failure of
graphene layer has been observed when temperature exceeds $3000$ K. Another sample
application for high-melting-point materials can be found in the extremely-high-temperature
plasma environment, such as fusion plasma. For example, in the record-holding Wendelstein
7-X fusion chamber the interior walls are cladded by graphite tiles and will be upgraded by
those of carbon fiber-reinforced carbon \cite{Wendelstein}. Also, recent progress has been
made on graphene fibers composed of different sizes of graphene sheets that have been
studied thermally and mechanically up to $3123$ K \cite{XinScience15}. Applications of
these graphene-based materials at even higher temperature are foreseeable, given e.g.,
the radiation from fusion plasma that can reach $2 \times 10^7$ $^\circ$C \cite{Wendelstein}.
All these applications require extraordinary mechanical properties of the material at
ultrahigh temperature with the brittle failure behavior being suppressed, reflecting
the significance of sample plasticity and ductility.

However, for the associated mechanical properties of 2D systems what
is lacking so far is a comprehensive study of the ductile behavior at high enough
temperature under mechanical deformation, particularly the possible brittle-to-ductile
transition that is important for material mechanical applications,
while the corresponding mechanisms still remain elusive for graphene-type 2D materials.
A key factor to be investigated here is the GB-controlled plasticity, which is crucial
for determining material properties such as strength, ductility, and deformation dynamics,
and for identifying the deformation mechanisms of polycrystalline graphene. The plastic
behavior is of particular importance for the high-temperature mechanical response,
but such studies for high-temperature graphene are still far from complete in the
existing research with a limited range of system temperature being examined. A main
obstacle here is the seemingly lack of mechanisms for ductile/plastic deformation
in 2D materials: While in traditional 3D bulk materials the brittle-to-ductile transition
and the occurrence of plastic deformation and flow are closely related to the dislocation
sources and new dislocation generation inside the system, in 2D graphene monolayers both
experiments \cite{LeeScience13} and atomistic simulations \cite{Grantab946,Wei12,
LIU20123465,WU20131421,YI2013373,CHEN2015135,Yang2015} showed a lack of dislocations
inside the grain interiors (other than being constrained at the GBs).

Most of computational studies of mechanical properties of graphene, primarily based on
atomistic simulations like molecular dynamics (MD), focused on low- or intermediate-temperature
behaviors which yield brittle fracture. For example, Grantab \textit{et al.} \cite{Grantab946}
investigated the role of GB misorientation angle on the fracture strength of graphene sheets
under uniaxial tension applied both perpendicular and parallel to the GB, and found that
larger angle leads to stronger GBs. Later studies by Wei \textit{et al.} \cite{Wei12} and
Liu \textit{et al.} \cite{LIU20123465} with tensile loads perpendicular to the GBs showed
that the local arrangement and distribution of GB dislocations is also an important factor
in determining the sample strength. In addition, atomistic details of failure and cracking
have been examined for GBs composed of $5|7$ dislocations. For example, the initiation of
sample failure was found to occur via the bond breaking in the heptagons of $7|6$ rings at
the GB \cite{Grantab946,WU20131421}. Recently, some MD simulations have also targeted the
important effect of temperature on the mechanical properties of graphene sheets. Results
from Yi \textit{et al.} \cite{YI2013373} showed a significant decrease of fracture strength
and strain as temperature increases (from 0 to 1800 K) when pulling along the direction
either parallel or perpendicular to the GB, which was attributed to stronger thermal
fluctuations and larger initial lengths of critical bonds at elevated temperatures. A similar
trend has been obtained in polycrystalline graphene \cite{CHEN2015135}. Interestingly,
recent MD simulations by Yang \textit{et al.} \cite{Yang2015} indicated a brittle-plastic
transition in nanocrystalline graphene at high enough temperatures (above 1000 K) and low
enough strain rate of the external tensile load, for which the appearance of plasticity was
linked to the degree of bond rotation and suppression of bond breaking at the GBs. However,
one of the key features of plastic deformation, the motion of dislocation defects, has not
been observed during the mechanical deformation of graphene across the temperature range of
existing studies.

Although the atomistic simulation techniques used in most studies of graphene, such as MD
and first-principles density functional theory (DFT), can well capture the microscopic
details of the system, they are restricted by not only the limited size of simulated samples,
but also the small accessible time range of evolution (given the atomic vibration scales that
are much shorter than the typical timescales of atomic diffusion and dislocation dynamics),
a factor that hinders the effective study of detailed process of plastic deformation. In recent
years, much progress has been made on the development of advanced techniques that can overcome
these limitations. Among them one of the fast-developing multiple-scale approaches is the phase
field crystal (PFC) modeling method \cite{PhysRevLett.88.245701,PhysRevE.70.051605,ElderPRB07,
PhysRevLett.96.225504,Mkhonta13}, as being used in this work. The PFC method has the advantage
of incorporating both microscopic crystalline structures and particle diffusive timescales,
and thus being able to effectively model large-scale material systems across an evolution time
regime well beyond that of atomistic methods. Another advantage of PFC is that the atomic
configurations of model system (including the defect core structures) form naturally during
the evolution of PFC atomic density field, without the need and limit of preconstructing
the detailed lattice and defect structures as in atomistic simulations. This is well suited
for the investigation of complex ordered structures and defect configurations, and the method
has been successfully applied to the study of a wide range of material systems
\cite{PhysRevLett.88.245701,PhysRevE.70.051605,ElderPRB07,PhysRevLett.96.225504,Mkhonta13,
PhysRevLett.118.255501,PhysRevE.80.046107,SalvalaglioPRMater18,SkaugenPRL18,GUO2018175,
ZHOU2017121,GAO2016238,0965-0393-24-5-055010} including the modeling of graphene
\cite{PhysRevB.94.035414,HirvonenSciRep17,SmirmanPRB17,LI201836,SeymourPRB16}.

In this study, we use the PFC model to investigate the deformation dynamics of various types
of symmetric tilt grain boundaries of graphene under uniaxial tensile tests, particularly
the phenomena of yield and plastic flow dominated by dislocation motion, which are missing
in previous work. A focus here is on mechanical behavior of graphene at ultrahigh temperatures
(beyond the range of most of previous studies) and small enough strain rate that is not
accessible to traditional atomistic simulations, as characterized by the dislocation-mediated
jerky plastic flow around the GBs under external stress. The related mechanisms of dislocation
dynamics during sample yielding, and of the brittle-to-ductile transition with the increase
of temperature, are also examined. Our findings include the stick-and-climb-glide type behavior
of individual dislocations that accounts for the occurrence of jerky plasticity beyond the
yield point, with the detailed dynamics (climb and glide) depending on the specific defect
structures. It is determined by the ability of existing dislocations to overcome the
pinning barriers after each intermittent waiting period of defect pinning and stress
accumulation, without the traditional mechanisms for the generation of new dislocations.
Our modeling approach is validated through the study of brittle fracture, which can still
occur at fairly high temperature. The outcomes are similar to those of previous
MD and experimental low-temperature studies showing crack initiation at the GB through the
formation of nanovoids and crack propagation into grain interiors, in addition to a new
type of failure dynamics associated with the migration of dislocation-bound nanovoids that
leads to low-angle GB disintegration. These results are expected to further our understanding
of the complex deformation mechanisms particularly the high-temperature mechanical response
of graphene-type 2D materials.

\section{Model and Methods}

\subsection{Phase field crystal model}
\label{sec:model}

In the PFC approach, which was introduced to efficiently model spatially periodic systems
with atomic-scale resolution through a continuum density field, the simplest dimensionless
free energy functional incorporating elasticity and plasticity in a single-component system
is expressed as \cite{PhysRevLett.88.245701,PhysRevE.70.051605}
\begin{equation}\label{1101}
  F=\int d\vec{r}\left\{\frac{1}{2}\phi \left [\epsilon+(\nabla+1)^2 \right ]\phi
  +\frac{\tau}{3}\phi^3+\frac{1}{4}\phi^4\right\},
\end{equation}
where $\phi$ is the atomic number density variation field, the coefficient $\tau$ can be connected
to three-point interparticle correlation \cite{HuangPRE10}, and $\epsilon$ is a phenomenological
temperature parameter controlling the degree of undercooling from the melting temperature, with
$\epsilon<0$ corresponding to a crystalline state and a higher (lower) magnitude $|\epsilon|$
corresponding to a lower (higher) temperature.
The free energy functional in Eq.~(\ref{1101}) can be derived from the classical density functional
theory (CDFT) of freezing, as has been discussed in detail elsewhere \cite{ElderPRB07,HuangPRE10}.
The corresponding equilibrium phase diagram has also been identified \cite{PhysRevE.70.051605},
giving the homogeneous and crystalline solid phases (in both 2D and 3D) and their coexistence,
as controlled by the temperature parameter $\epsilon$ and the average density variation $\phi_0$.
In 2D systems, when $\tau+3\phi_0>0$ with large enough $\phi_0$ the corresponding PFC solid state
is of honeycomb lattice symmetry. Such a model has been successfully used to investigate the
properties of GBs \cite{PhysRevB.94.035414}, triple junctions \cite{HirvonenSciRep17},
Moir\'{e} patterns \cite{SmirmanPRB17}, and polycrystals \cite{PhysRevB.94.035414,LI201836}
of 2D graphene monolayers, and is adopted in this work for further study of graphene
defect dynamics. We have parameterized this PFC model for graphene, with the detailed process
presented in \ref{sec:parameterization}, including matching parameter $\epsilon$ to real
temperature scale and identifying the conversion from PFC to real units for stress, elastic
modulus, and length.

The standard PFC evolution equation governing the conserved dynamics of density field is described by
\begin{equation}\label{1106}
  \frac{\partial\phi}{\partial t}=\nabla^2 \frac{\delta F}{\delta\phi}
  =\nabla^2 \left [\epsilon\phi+(1+\nabla^2)^2\phi+\tau\phi^2+\phi^3 \right ],
\end{equation}
such that the system, with its PFC-characterized atomic-scale microstructure, evolves in a
diffusive timescale. Thus this PFC model can effectively capture the slow, diffusive dynamic
processes during e.g., nucleation, solidification, and structural transformation of material
systems, but not the processes involving collective atomic motion or phonon-level dynamics
which occurs on much faster timescales. It then cannot fully describe the mechanisms of
mechanical response of crystalline solids with elastic and plastic relaxation.
To tackle this problem, very recently we developed a computational scheme based on an interpolation
algorithm for PFC (IPFC) \cite{PhysRevE99013302}, to effectively simulate the process of mechanical
relaxation in the PFC model. The method was applied to the study of mechanical properties and brittle
fracture behavior of a single-crystal nanoribbon under uniaxial tensile test \cite{PhysRevE99013302},
with results qualitatively consistent with those of previous atomistic calculations of pristine
graphene (by e.g., MD \cite{ZhaoNanoLett09,doi:10.1063/1.3488620}, Monte Carlo \cite{ZakharchenkoPRL09}
and \textit{ab initio} DFT \cite{LiuPRB07}). Here this IPFC scheme is adopted to examine the
mechanical behavior of graphene GBs under uniaxial tension, with details of the simulation
system setup given below.

\subsection{System setup}

\begin{figure}
\centerline{
\includegraphics[width=0.35\textwidth]{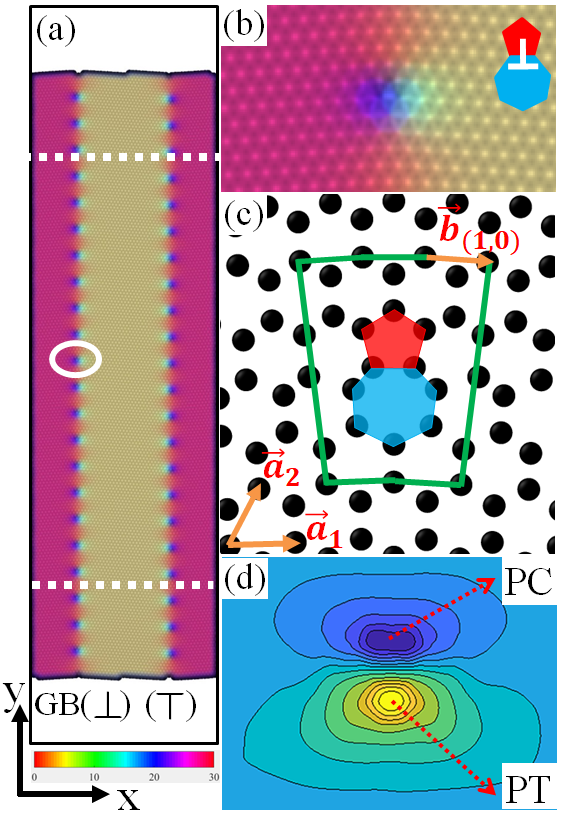}}
  \caption{Structure of a simulated bicrystal containing armchair symmetric tilt grain boundaries
    (with tilt angle $\theta=7.34^\circ$). (a) Illustration of lattice orientation in the simulated
    system, where different colors correspond to different local crystallographic orientations
    (with color bar given at the bottom). The two dashed lines indicate the boundaries of the
    active zone. (b) Zoomed-in view of the circled region in (a). (c) The corresponding atomic
    structure of the $5|7$ dislocation in (b) and its Burgers vector $\vec b=(1,0)$. (d) Contour
    of the local strain $\varepsilon_{yy}$ distribution for (c), with the locations of peak
    compression (PC) and peak tension (PT) indicated.}
  \label{fig:GB_structure}
\end{figure}

The simulations are conducted for a 2D PFC system scaled by $L_x\times L_y$ (chosen as
$512\Delta x \times 2048\Delta y$ in this work, corresponding to $13.6$ nm $\times$ $54.5$ nm)
with periodic boundary condition in both directions, as illustrated in Fig.~\ref{fig:GB_structure}(a)
with the use of lattice orientation contour \cite{WANG20132489}. The simulation box consists of
three parts, with the first two being the solid crystalline regions coexisting with a third part
of homogeneous phase (white margins of thickness $200\Delta y$ at each of the two vertical ends).
Within the two solid regions that are stretched along the $y$ direction, the first part is set as
the active zone (i.e., the middle area between the white dashed lines in Fig.~\ref{fig:GB_structure}(a),
of initial size $L_x \times L_y^0 = 512\Delta x \times 1248\Delta y$ or $13.6$ nm $\times$ $33.2$ nm),
in which the atomic structure mechanically deforms and relaxes following the external tensile load;
the second part is the traction zones located at both ends of the solid where the loading is imposed
(with fixed size $L_x \times L_{yt} = 512\Delta x \times 200\Delta y$ at each side). The atomic density
field in the traction zones is fixed and predetermined as $\phi_{\rm trac}$, which is the density
profile of the corresponding regions when the whole system reaches equilibrium before the tensile
deformation is applied. These two traction regions will then be moved adiabatically along the
direction of tensile loading, while the system is subjected to the traction boundary condition
\cite{PhysRevE.80.046107} for which the free energy functional Eq.~(\ref{1101}) is modified by
$F \rightarrow F + F_{\rm ext}$, with $F_{\rm ext}=\int d\vec r M(\vec r) [\phi(\vec r)-\phi_{\rm trac}(\vec r)]^2$
serving as free energy penalty. This penalty term is zero inside the active zone and the homogeneous
regime (by setting $M(\vec r)=0$), and is always positive in the traction regions (by setting e.g.,
$M(\vec r)=2$) so that the associated density field $\phi_{\rm trac}$ can be imposed.

When the solid sheet is pulled at two ends along the $y$ direction, the length of the active zone is
increased by $2\Delta y$ at each step. An interpolated scheme of the PFC model (i.e., the IPFC scheme
\cite{PhysRevE99013302}) is used to enable fast mechanical relaxation of the deformed system. At each
numerical grid position it first estimates the updated value of the density field $\phi$ as a result of
current-step mechanical deformation through a linear interpolation of the previous-step old density
field, based on the assumption of local instantaneous elastic equilibration and the linear displacement
in the limit of small strain increment; the system is then relaxed through PFC dynamics (i.e.,
Eq.~(\ref{1106})) to reach mechanical equilibrium before the next-step tensile deformation. More
details of the algorithm as well as the way of evaluating engineering strain $\varepsilon_e$ and
stress $\sigma_e$ during the deformation process are given in Ref.~\cite{PhysRevE99013302}.

It is noted that the PFC model systems examined here are restricted to purely 2D, which would
correspond to the scenario of epitaxial overlayer confined by the substrate such that effects
of out-of-plane deformation are neglected. On the other hand, in real 3D systems out-of-plane
corrugations should still have impacts on material properties, particularly for small-angle GBs
(while such buckling would be significantly reduced for large GB misorientation angles with high
dislocation density and/or by binding to the substrate) \cite{PhysRevB.81.195420,doi:10.1021/nl100988r}.
For the case of tensile loading studied here, effects of any vertical variations are mitigated
by the imposing of lateral stretching which tends to flatten the sheet. Thus the initial structure
of out-of-plane deformation would be of less influence at larger applied strain, and is not
expected to play a significant role (other than some small quantitative variations) on the
behavior of plastic regime which is the focus of this work. It also would not affect the basic
mechanisms of yielding or sample failure that will be identified below, particularly for any
possible motion of dislocations which should still be mainly of 2D nature in a stretched epitaxial
monolayer.

To generate the results of GBs, the PFC dynamic equation (\ref{1106}) is solved numerically
through a pseudospectral method. Details of the numerical scheme are given in \ref{sec:numerical}.
Note that the PFC model equation studied here is deterministic, without noise dynamics.
The only possible randomness involved is in the initial condition setup. Our calculations
show that different setup of initial noise in the system would not change the
steady-state GB results that are used for the subsequent mechanical study, as the system
always evolves to the same minimum-energy steady state; thus it would not affect the mechanical
behavior and deformation properties of the system due to the deterministic PFC and IPFC
dynamics used. In this work, unless otherwise specified, the model parameters used in the
simulations are chosen as $\tau=1$, the numerical grid spacing $\Delta x=\Delta y=\pi/4$,
time step $\Delta t=0.2$, and the strain rate $\dot{\varepsilon}_e=8.01 \times 10^{-7}$, with
the temperature parameter $\epsilon$ ranging from $-0.5$ to $-0.1$ (which corresponds to a
temperature range from $3352$ K to $4278$ K based on the model parameterization described
in \ref{sec:parameterization}).

\section{Results and Discussion}

\subsection{Structure and strain distribution of GBs}
\label{sec:GBs_structure}

\begin{figure*}
\centerline{
\includegraphics[width=0.48\textwidth]{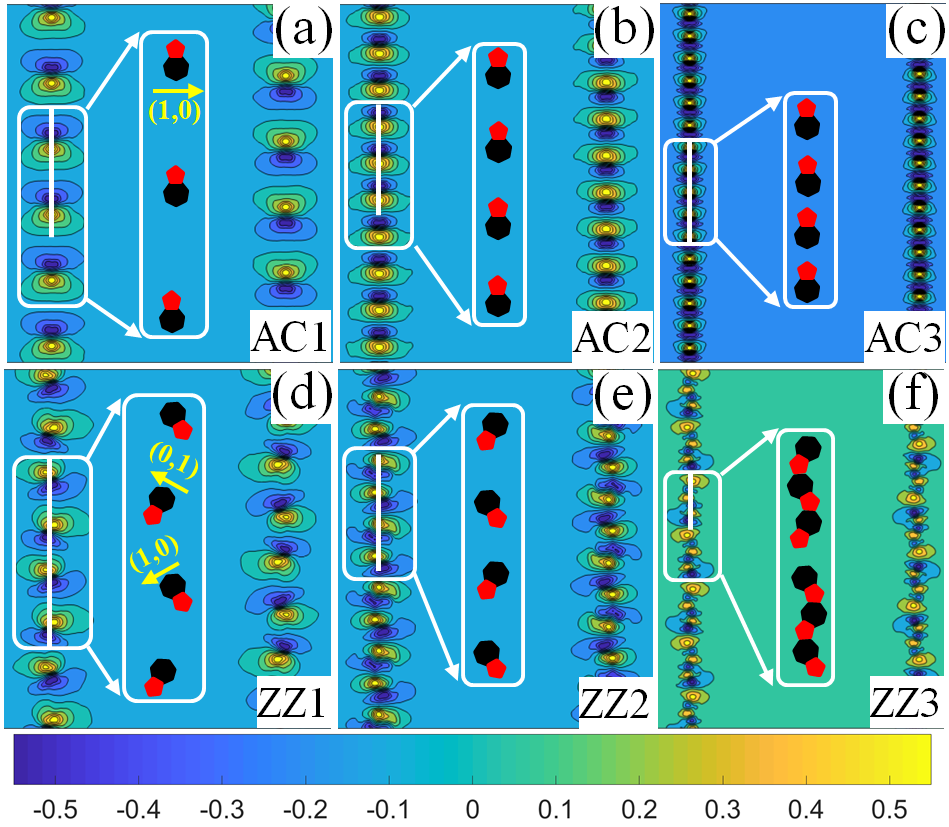}
\includegraphics[width=0.465\textwidth]{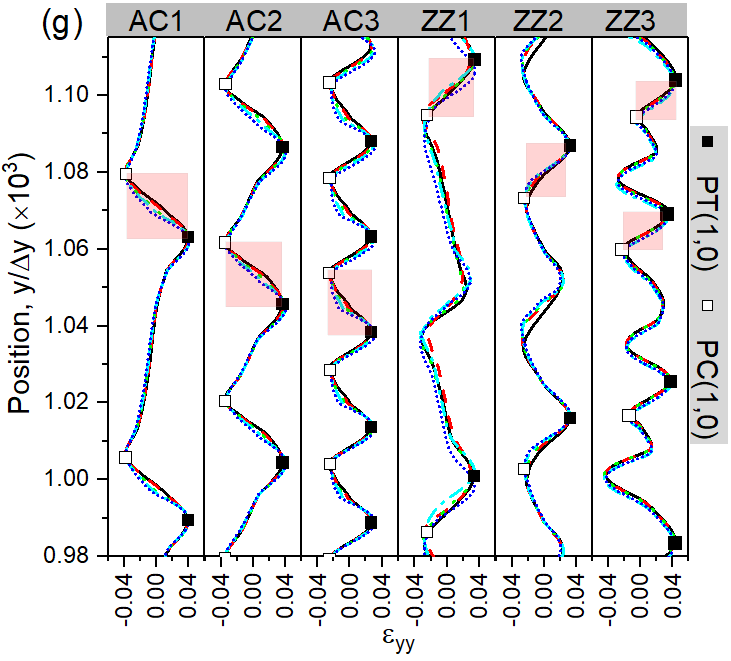}}
  \caption{Spatial distribution of local strain $\varepsilon_{yy}$ for different graphene GBs,
    including armchair GBs with $\theta = 7.34^\circ$ (denoted by AC1 in (a)), $13.17^\circ$
    (AC2 in (b)), and $21.79^\circ$ (AC3 in (c)), and zigzag GBs with $\theta = 8.61^\circ$
    (ZZ1 in (d)), $13.17^\circ$ (ZZ2 in (e)), and $23.49^\circ$ (ZZ3 in (f)). Schematics of
    dislocations composed of pentagons (red) and heptagons (black) in the white boxed regions
    are shown in the middle of each panel. The cross-section profiles of $\varepsilon_{yy}$
    along the GB line (corresponding to the white vertical lines in (a)--(f)) are given in
    (g), for different values of temperature parameter $\epsilon=-0.5$ ($T=3352$ K; solid
    black curves), $-0.4$ ($3584$ K; dashed red), $-0.3$ ($3815$ K; dotted green), $-0.2$
    ($4047$ K; dot-dashed cyan), and $-0.1$ ($4278$ K; dotted blue).}
  \label{fig:GBs}
\end{figure*}

In most of our simulations the system is set up as a bicrystal containing two parallel symmetric
GBs, each of misorientation angle $\theta$. Each GB separates two single crystalline graphene of
orientations $\theta/2$ and $-\theta/2$, with the boundary line along the $y$ direction (i.e.,
the direction of tensile loading), as shown in Fig.~\ref{fig:GB_structure}(a). No preconstruction
of detailed defect core structures and atomic configuration of GBs is needed in our PFC simulations;
instead, the initial condition at the location of each GB is set up as a narrow vertical strip of
supersaturated homogeneous phase connecting two misoriented single-crystalline regions. The PFC
density of each homogeneous strip is of the same value as the average density of the solid regions,
with a strip width of $20 \Delta x$. We have varied the strip width from $10 \Delta x$ to
$20 \Delta x$ and added different degree of initial noise in the system, all giving very similar
results. The subsequent dynamic relaxation of the PFC equation leads to spontaneous solidification
of this supersaturated strip which merges the two adjacent solid regions and forms a GB. The
system then evolves to a steady state with minimum free energy, resulting in the stable GB
configurations presented in Figs.~\ref{fig:GB_structure} and \ref{fig:GBs}. These GBs
generated from the evolution of PFC density field consist of arrays of $5|7$ (penta-hepta)
dislocation cores, consistent with the GB defect structure found in experiments and previous
atomistic simulations of graphene. An example of $5|7$ disperse dislocation in a small-angle
GB is given in Fig.~\ref{fig:GB_structure}(b)-(d), with Burgers vector $\vec b=(1,0)$
(based on the definition in Ref.~\cite{PhysRevB.81.195420}). Fig.~\ref{fig:GB_structure}(d)
shows the contour of the corresponding local strain $\varepsilon_{yy}=\partial u_y/\partial y$,
where $u_y$ is the $y$ component of the displacement field, as obtained through the method
given in Refs.~\cite{1s1s} and \cite{doi:10.1080}. Regions of compressive vs. tensile
strains can be identified for the two disclinations (5- vs. 7-membered rings), as
well as the locations of peak compression (PC) and peak tension (PT) as indicated in
Fig.~\ref{fig:GB_structure}(d). 
The midpoint between PC and PT can then be used to locate the exact position of the $5|7$
dislocation core.

Here we examine the detailed dynamics of low-, intermediate-, and high-angle
symmetric tilt GBs subjected to the tensile loading parallel to the GB line. These include
(i) armchair tilt GBs with misorientation angles $\theta = 7.34^\circ$ (denoted by AC1),
$13.17^\circ$ (AC2), and $21.79^\circ$ (AC3), where $\theta$ is measured between the armchair
edges of the two adjoining grains, and (ii) zigzag tilt GBs with $\theta = 8.61^\circ$ (ZZ1),
$13.17^\circ$ (ZZ2), and $23.49^\circ$ (ZZ3), as measured between the two zigzag orientations
of the grains. The corresponding GB structures and strain ($\varepsilon_{yy}$) contours before
the tensile loading are presented in Fig.~\ref{fig:GBs}(a)--(f), showing an increase of
dislocation density with larger $\theta$, as expected. For AC sheets, the GBs are composed of
a periodic array of $5|7$ isolated dislocations with identical Burgers vector $\vec b =(1,0)$,
and the spacing between neighboring $(1,0)$ dislocations decreases as $\theta$ increases (see
Fig.~\ref{fig:GBs}(a)--(c)). For AC3 with large tilt angle $\theta = 21.79^\circ$, there is
a separation of only one atomic spacing between any pair of disperse $(1,0)$ dislocations,
agreeing with the previous MD simulation results \cite{Grantab946,Wei12,LIU20112306}.
For zigzag sheets ZZ1 and ZZ2 at small and intermediate angles, each GB is constituted of
alternating $(1,0)$ and $(0,1)$ dislocations whose Burgers vectors vary by $60^\circ$ in
orientation (see Fig.~\ref{fig:GBs}(d), (e)). For ZZ3 with larger misorientation, each GB
consists of a series of alternating $(1,0)+(0,1)+(1,0)$ and $(0,1)+(1,0)+(0,1)$ dislocation
triplets (i.e., 3 connected $5|7$ dislocations), as shown in Fig.~\ref{fig:GBs}(f). The two
neighboring triplets are separated by exactly one hexagon ring. A brief summary of these GBs
examined in this work is given in Table \ref{table:GBs} of \ref{sec:parameterization}.

To further illustrate the spatial variation of strain around the dislocations, the one-dimensional
cross-section profiles of $\varepsilon_{yy}$ along the GB line are plotted in Fig.~\ref{fig:GBs}(g),
for all six armchair and zigzag GBs at different temperatures. The locations of strain peaks (PC
vs. PT) are indicated, corresponding to the penta-hepta pairs of disclinations (see e.g., the shaded
regions in Fig.~\ref{fig:GBs}(g)). As the misorientation angle increases from AC1 to AC3 or from ZZ1
to ZZ3, the distance between neighboring dislocation cores decreases due to higher dislocation density;
however, for AC GBs (AC1 to AC3) the distance between the two constituent disclinations of each
dislocation (i.e., between PC and PT positions of the shaded region), which is proportional to the
size of dislocation core region, remains almost unchanged, although the corresponding magnitude of
strain variation along the GB direction becomes smaller. By contrast, for ZZ GBs as $\theta$
increases (ZZ1 to ZZ3) both the dislocation core size and strain variation magnitude decrease.
This difference is attributed to the fact that different from the AC GBs that are characterized
by arrays of a single type of $(1,0)$ dislocations, the dislocations in ZZ GBs are of mixed type
(i.e., $(1,0)$ and $(0,1)$), leading to different degree of elastic interaction between dislocations
along the GB line particularly the cancellation of stress fields between nearby or connected
dislocations \cite{doi:10.1021/nn200832y,C6RA07584C}. Similar results of defect structures and
strain distribution in the absence of external deformation are obtained at different temperatures
(see e.g., close-to-overlapped plots in Fig.~\ref{fig:GBs}(g) for different temperature values).
They play an important role on determining the mechanical properties of GBs samples under external
loading, as will be detailed below.

\subsection{Effects of uniaxial tension: Brittle failure behavior}
\label{sec:mech_brittle}

\begin{figure}
\centering
\includegraphics[width=0.49\textwidth]{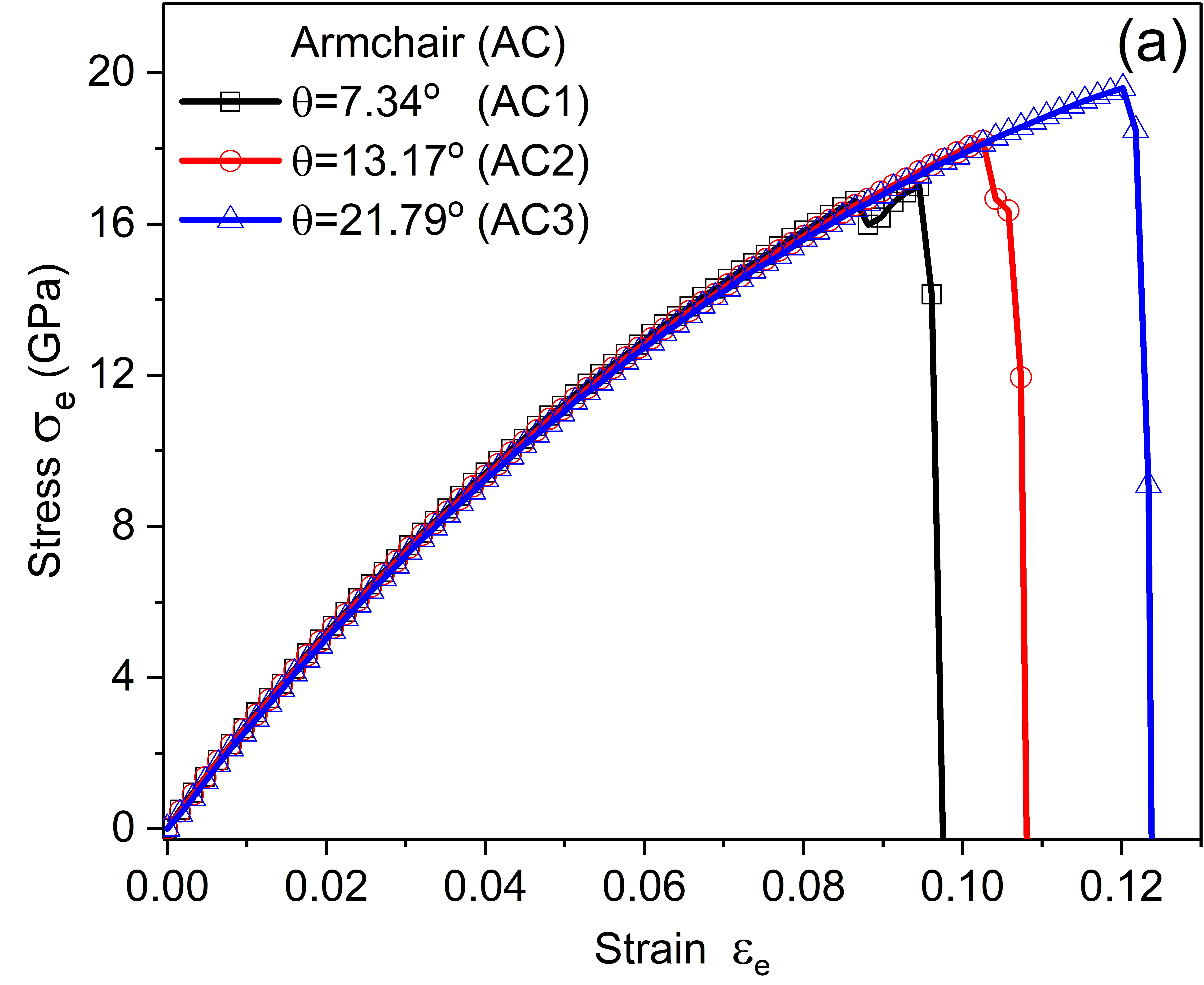}
\includegraphics[width=0.49\textwidth]{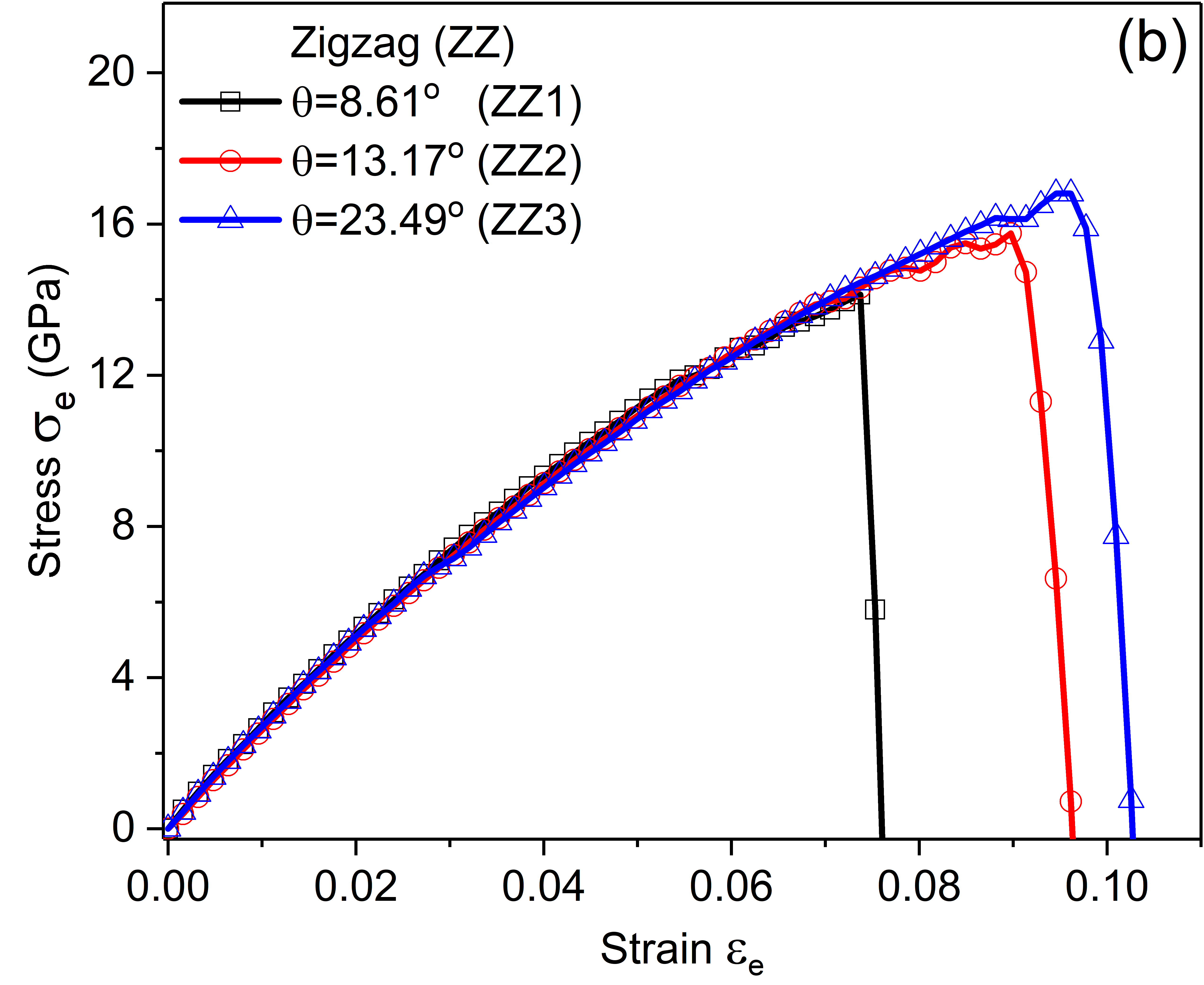}
\caption{Stress-strain curves for (a) AC and (b) ZZ sheets at $\epsilon=-0.5$ ($T=3352$ K) under
  uniaxial tension applied parallel to the GBs.}
  \label{fig:stress-strain}
\end{figure}

\begin{figure}
\centering
\includegraphics[width=0.485\textwidth]{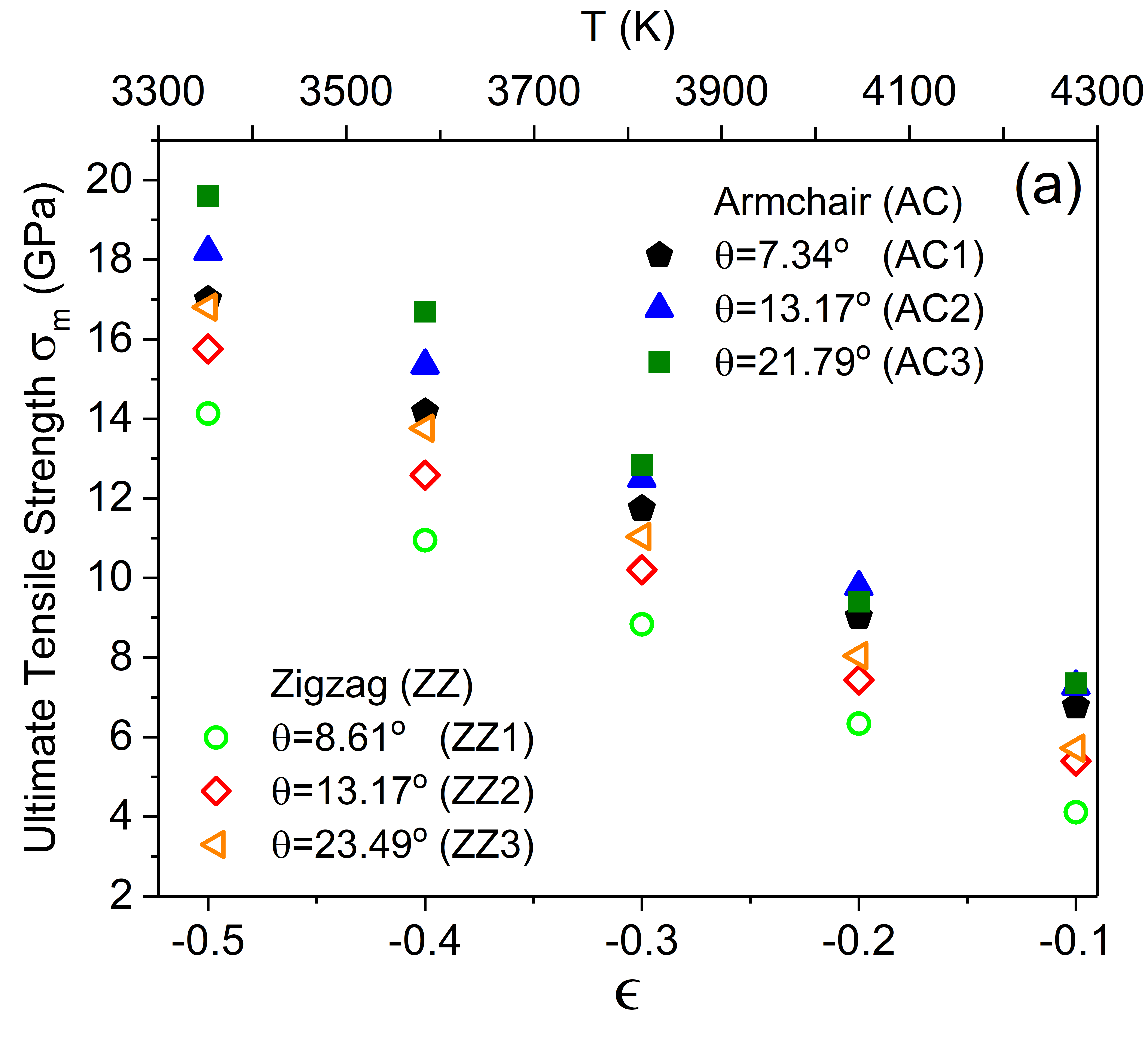}
\includegraphics[width=0.49\textwidth]{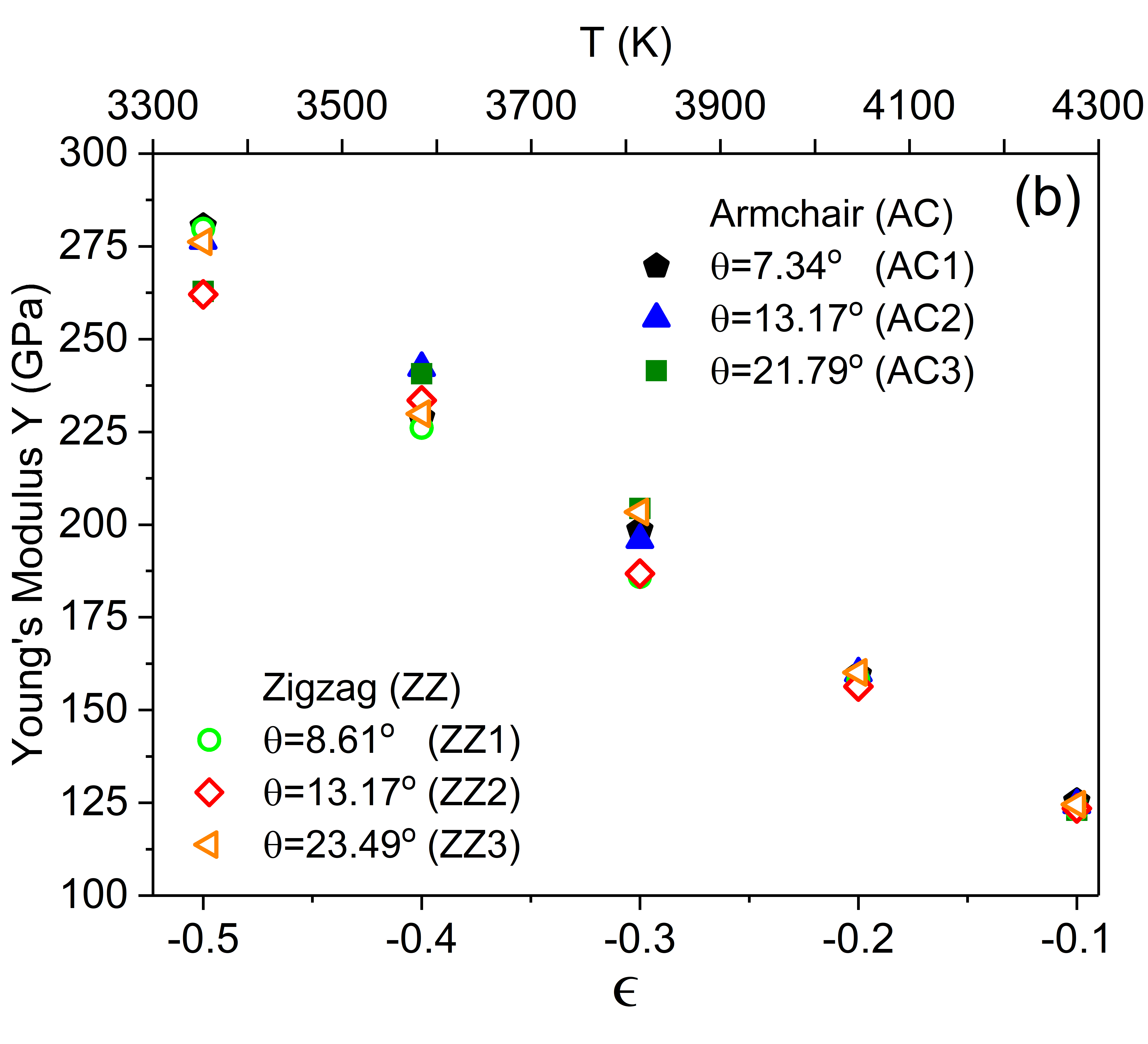}
  \caption{(a) Ultimate tensile strength and (b) Young's modulus as a function of temperature
    for various AC and ZZ GBs.}
  \label{fig:sigma_m_Y}
\end{figure}

\begin{figure}
\centering
\includegraphics[width=0.55\textwidth]{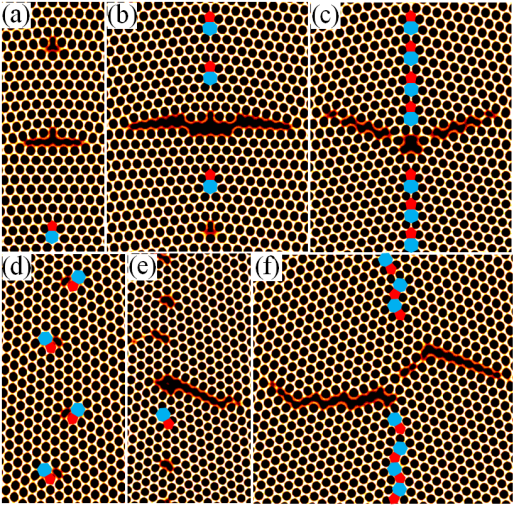}
\caption{Atomic structures at the stage of sample failure in AC and ZZ sheets at $\epsilon=-0.5$
  ($T=3352$ K), for (a) AC1 at applied engineering strain $\varepsilon_e=9.78\%$,
  (b) AC2 at $10.98\%$, (c) AC3 at $12.42\%$, (d) ZZ1 at $7.05\%$, (e) ZZ2 at $9.38\%$, and
  (f) ZZ3 at $10.26\%$.}
\label{fig:fracture_e05}
\end{figure}

Figures \ref{fig:stress-strain}--\ref{fig:fracture_e05} show the results of deformation or fracture
behavior and mechanical properties of these armchair and zigzag GBs after the imposing of uniaxial
tensile loads (with direction parallel to the GBs), which are consistent with those found in
previous experiments and MD simulations. An example of the stress-strain relation at $\epsilon=-0.5$
(corresponding to $T=3352$ K) is presented in Fig.~\ref{fig:stress-strain}, for both AC and ZZ GBs
under uniaxial tension. A brittle-type failure behavior is found for these symmetric GBs of different
misorientation angles, similar to previous MD results for graphene calculated at low (or $0$ K)
temperature with loading direction either parallel or perpendicular to the GB
\cite{Grantab946,Wei12,LIU20123465}, and also to the corresponding PFC calculations conducted
at a low temperature of $111$ K ($\epsilon=-1.9$) in \ref{sec:numerical}.

In both cases of AC and ZZ GBs with regularly distributed $5|7$ dislocations studied here,
larger tilt angles $\theta$ correspond to stronger GBs with larger strain value at failure
and higher ultimate strength (defined as the maximum tensile stress during deformation;
see Fig.~\ref{fig:stress-strain}(a), (b) and Fig.~\ref{fig:sigma_m_Y}(a) at high temperatures
and also Fig.~\ref{fig:lowT} in \ref{sec:numerical} at a low temperature), which
qualitatively agrees with previous MD \cite{Grantab946,Wei12,LIU20123465}
and experimental \cite{RasoolNatCommun13} findings. This can be attributed to higher density
of dislocation arrays at larger $\theta$, which leads to a larger degree of mutual cancellation
of dislocation stress fields and thus smaller prestrain \cite{doi:10.1021/nn200832y}, yielding
larger failure strain and GB strength. More quantitative details will be given below.
On the other hand, as shown in Fig.~\ref{fig:sigma_m_Y}(b) values of Young's modulus $Y$,
as calculated from the small-strain linear part of the stress-strain curves, depend weakly on the
misorientation angle and are close to that of the pristine sample (e.g., at $\epsilon=-0.5$
($T=3352$ K) $Y$ values of various GBs are close to the value of $284$ GPa obtained for the
corresponding armchair single crystal of the same size). This is consistent with experimental
measurements showing very limited influence of grain boundaries on the elastic moduli of graphene
\cite{LeeScience13,doi:10.1021/nl200429f}. We note that values of stress and Young's modulus
calculated here are much smaller than those reported in previous MD simulations, which is
due to the much higher temperature range examined in this work. If using low temperatures
(see Fig.~\ref{fig:lowT} at $T=111$ K in \ref{sec:numerical} for both AC and ZZ GBs and also
Fig.~\ref{fig:parameterization} in \ref{sec:parameterization} for pristine graphene), the
values obtained from our PFC modeling are comparable to atomistic simulation results.

Snapshots of the corresponding atomic structures during deformation and fracture are given in
Fig.~\ref{fig:fracture_e05}, for both AC and ZZ sheets. Our PFC simulations show that failure
or cracking of the GB sheet originates from the breaking of rings at or next to $5|7$
dislocations and hence the formation of nanovoids at the GB, before the cracks propagate
into the interior of the grain. This scenario well agrees with experimental observation
\cite{KimNanoLett12,RasoolNatCommun13} showing crack propagation inside the grain following
the armchair or zigzag lattice direction but not along the GB line, as seen in our
results of Fig.~\ref{fig:fracture_e05} and also Fig.~\ref{fig:lowT} in \ref{sec:numerical}.
Our finding also agrees with previous atomistic simulations of graphene where the
bond breaking at the defect rings of GBs was observed in MD as the
initiation of sample failure \cite{Grantab946,WU20131421}, and the increase
of ultimate strength with the GB tilt angle has been attributed to the reduction of initial
prestrain within the defect rings or bonds as the angle increases, based on MD and quantum
DFT calculations \cite{Grantab946}. As indicated in Fig.~\ref{fig:GBs}(g) and discussed in
Sec.~\ref{sec:GBs_structure}, our PFC results give a similar trend of preexisting strain
around dislocation cores before any tensile deformation is applied: For AC sheets with GB
angle $\theta$ increasing from $7.34^\circ$ (AC1), $13.17^\circ$ (AC2), to $21.79^\circ$ (AC3),
the strain amplitude between the penta-hepta disclination pair is measured as 7.93\%, 6.92\%,
and 5.21\% respectively; Similarly, for ZZ sheets the strain amplitude estimated in the regions
of Fig.~\ref{fig:GBs}(g) decreases from 5.91\%, 5.74\%, to 5.6\% when $\theta$ increases from
$8.61^\circ$ (ZZ1), $13.17^\circ$ (ZZ2), to $23.49^\circ$ (ZZ3). This decrease of preexisting strain
caused by dislocation defect rings results in the enhancing of GB strength and larger failure
strain with higher tilt angle, as seen in Figs.~\ref{fig:stress-strain}, \ref{fig:sigma_m_Y}(a),
and \ref{fig:lowT} and consistent with MD results for graphene \cite{Grantab946}.

Interestingly, a new and different phenomenon of deformation dynamics is observed for ZZ GBs
at low enough angle, e.g., ZZ1 sheet with $\theta=8.61^\circ$, as compared to the intermediate-
and high-angle ZZ GBs (ZZ2 and ZZ3 in Fig.~\ref{fig:fracture_e05}(e) and (f)). As shown in
Fig.~\ref{fig:fracture_e05}(d), for this low-angle GB characterized by well-separated $(1,0)$
and $(0,1)$ dislocations, although the failure still initiates around the GB $5|7$ dislocations
the subsequent deformation dynamics is featured by the motion (mostly glide) of each
individual dislocation bound together with its connected nanovoid. The glide directions of
$(1,0)$ and $(0,1)$ dislocations are opposite to each other, both perpendicular to the external
pulling direction, leading to the splitting of the GB dislocation array into two sub-arrays of
$(1,0)$ and $(0,1)$ types respectively. It is noted that although these disperse $(1,0)$ and
$(0,1)$ dislocations have been examined in MD \cite{C6RA07584C,doi:10.1021/nl100988r},
first-principles DFT \cite{PhysRevB.81.195420,ZHANG2013151}, and 2D PFC \cite{PhysRevB.94.035414,
LI201836} studies, the focus there was on the corresponding GB energies without
external stress. For the case of tensile deformation and fracture, in previous
MD simulations of low enough tilt angles \cite{Grantab946,Wei12,LIU20123465,YI2013373} the ZZ
GBs with $(1,0)+(0,1)$ paired dislocations (i.e., the bound $5|7|5|7$ pairs) were studied instead,
which yields the normal behavior of cracking that initiates at and propagates from the defect
locations without the motion of dislocations (similar to that of AC and ZZ2, ZZ3 GBs shown in
Fig.~\ref{fig:fracture_e05}), a scenario different from the failure dynamics of GB disintegration
observed here in the small-angle ZZ1 sheet. For 2D planar systems as examined in this work,
it has been found by DFT \cite{PhysRevB.81.195420} and combined MD and PFC \cite{LI201836}
calculations that without external deformation, the lowest-energy structures of intermediate-
and low-angle ZZ GBs are those of $(1,0)$ and $(0,1)$ disperse $5|7$ dislocations, instead of
the paired ones, consistent with our finding here. Even when out-of-plane deformations are
considered, recent results combining MD and Read-Shockley-type dislocation model indicated
that this disperse $(1,0)$ and $(0,1)$ structure is also the lowest-energy state of ZZ GBs
at small enough angles (e.g., $\theta \simeq 10^\circ$) \cite{C6RA07584C}.

\subsection{Effects of uniaxial tension: Plastic flow at ultrahigh temperature}
\label{sec:mech_highT}

We have further investigated the influence of temperature on the mechanical properties and deformation
dynamics of graphene grain boundaries. Fig.~\ref{fig:sigma_m_Y} shows the variations of the ultimate
tensile strength $\sigma_m$ and Young's modulus $Y$ with respect to temperature (ranging between
$\epsilon=-0.5$ ($T=3352$ K) and $-0.1$ ($4278$ K)) for both AC and ZZ GBs. Clearly, with the
increase of temperature the ultimate strength $\sigma_m$ of AC and ZZ sheets decreases monotonously
(Fig.~\ref{fig:sigma_m_Y}(a)), similar to the temperature dependence of the failure strength of
graphene GBs obtained in MD \cite{YI2013373} although here the range of temperatures examined is
much higher. In addition, the result of larger $\sigma_m$ at higher GB angle $\theta$ holds for
different temperatures given the effect of prestrain described above, although the variation becomes
smaller at higher temperature. For Young's modulus $Y$, it is found to also decrease with the increase
of temperature for both AC and ZZ sheets (Fig.~\ref{fig:sigma_m_Y}(b)), consistent with the MD results
for pristine graphene \cite{doi:10.1063/1.3488620}. Across the temperature range examined in this
study, only weak dependence of $Y$ on the tilt angle $\theta$ is obtained, especially for ultrahigh
temperatures.

\begin{figure*}
\centering
\includegraphics[width=0.8\textwidth]{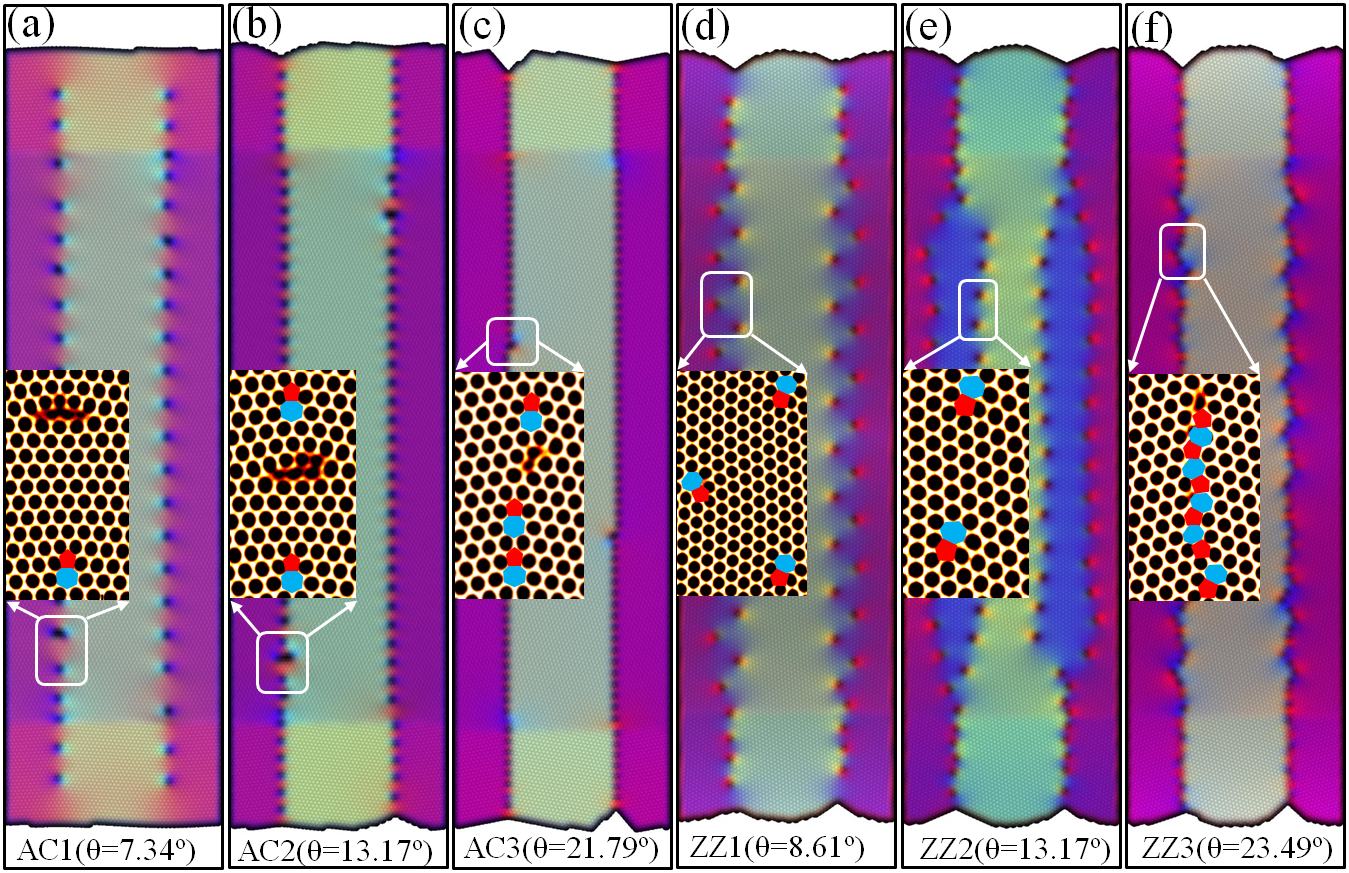}
\caption{Spatial distribution of local lattice orientation at ultrahigh temperature ($\epsilon=-0.1$
  with $T=4278$ K), for (a) AC1 at $\varepsilon_e=8.56\%$, (b) AC2 at $9.60\%$, (c) AC3 at $10.07\%$,
  (d) ZZ1 at $5.45\%$, (e) ZZ2 at $7.13\%$, and (f) ZZ3 at $7.18\%$. The atomic structure of the
  boxed region is shown inside each panel.}
\label{fig:deformation_e01}
\end{figure*}

\begin{figure}
\centering
\includegraphics[width=0.7\textwidth]{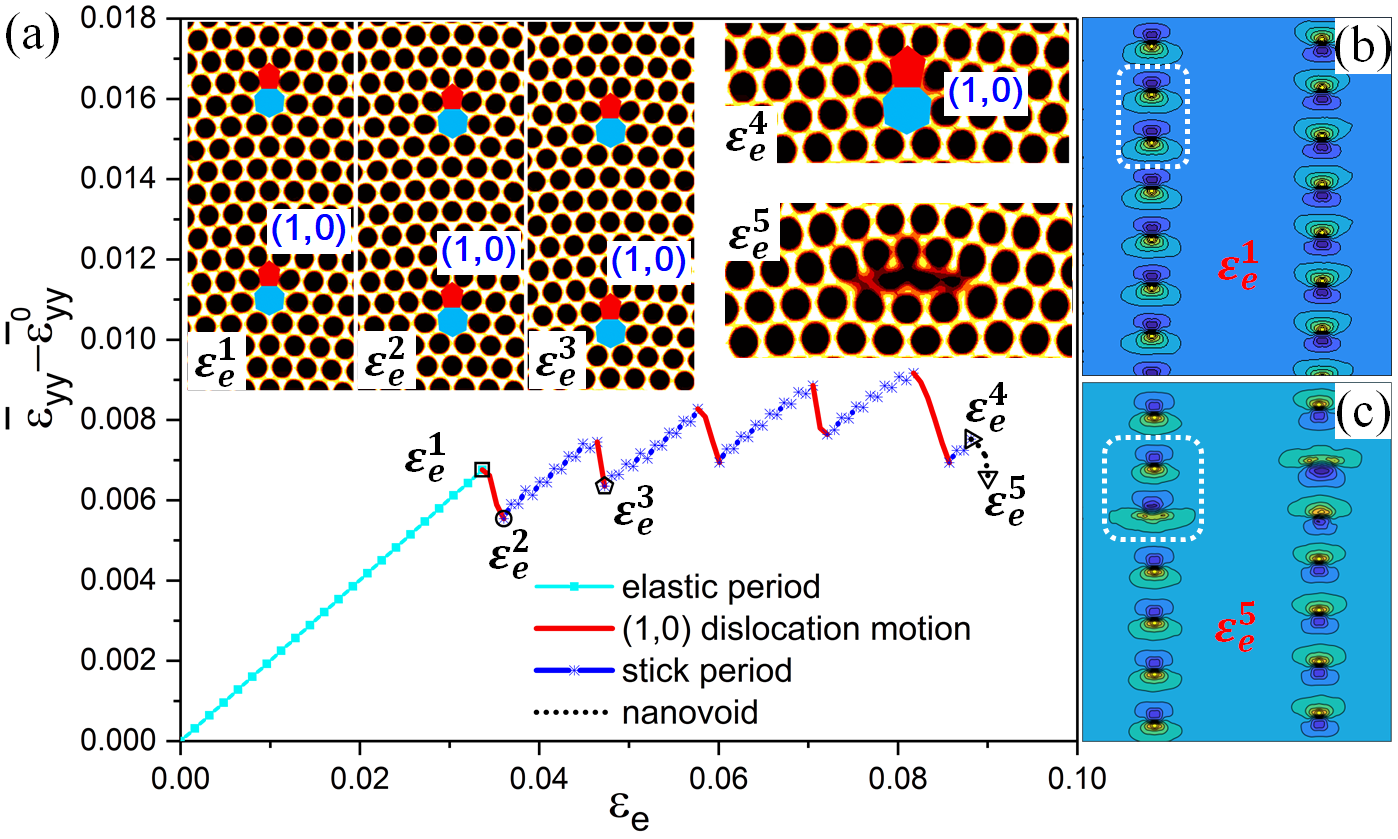}
\caption{(a) Change of average local strain, $\bar{\varepsilon}_{yy}-\bar{\varepsilon}_{yy}^0$,
  as a function of applied strain $\varepsilon_e$ for AC1 sheet at $\epsilon=-0.1$ ($T=4278$ K).
  The insets show the dynamic process of dislocation motion and the formation of nanovoid,
  corresponding to snapshots at $\varepsilon_e^1 = 3.37\%$, $\varepsilon_e^2 = 3.61\%$,
  $\varepsilon_e^3 = 4.73\%$, $\varepsilon_e^4 = 8.81\%$, and $\varepsilon_e^5 = 8.99\%$.
  (b) and (c) give the contours of the local strain $\varepsilon_{yy}$ distribution at
  $\varepsilon_e^1$ and $\varepsilon_e^5$ respectively, with the boxed region corresponding
  to the two dislocations shown in the insets of (a).}
  \label{fig:AC1_e01}
\end{figure}

\begin{figure}
  \centerline{
    \includegraphics[width=0.7\textwidth]{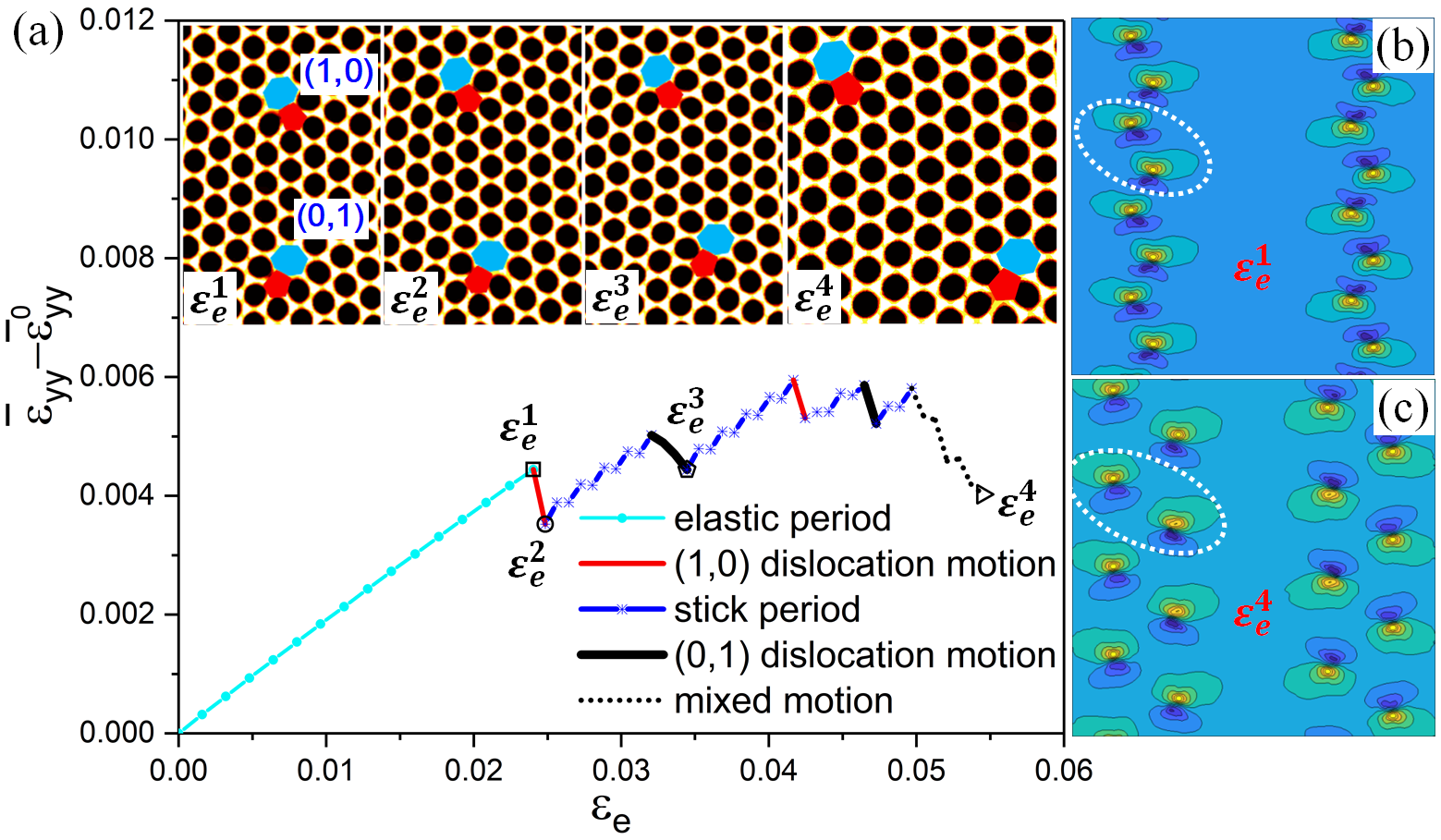}}
  \caption{(a) Change of average local strain, $\bar{\varepsilon}_{yy}-\bar{\varepsilon}_{yy}^0$,
    as a function of applied strain $\varepsilon_e$ for ZZ1 sheet at $\epsilon=-0.1$ ($T=4278$ K).
    The insets show the dynamic process of dislocation migration, corresponding to snapshots at
    $\varepsilon_e^1 = 2.40\%$, $\varepsilon_e^2 = 2.48\%$, $\varepsilon_e^3 = 3.45\%$, and
    $\varepsilon_e^4 = 5.45\%$. (b) and (c) give the contours of the local strain $\varepsilon_{yy}$
    distribution at $\varepsilon_e^1$ and $\varepsilon_e^4$ respectively, with the boxed region
    corresponding to the two dislocations shown in the insets of (a).}
  \label{fig:ZZ1_e01}
\end{figure}

\begin{figure*}
  \centering
  \includegraphics[width=0.63\textwidth]{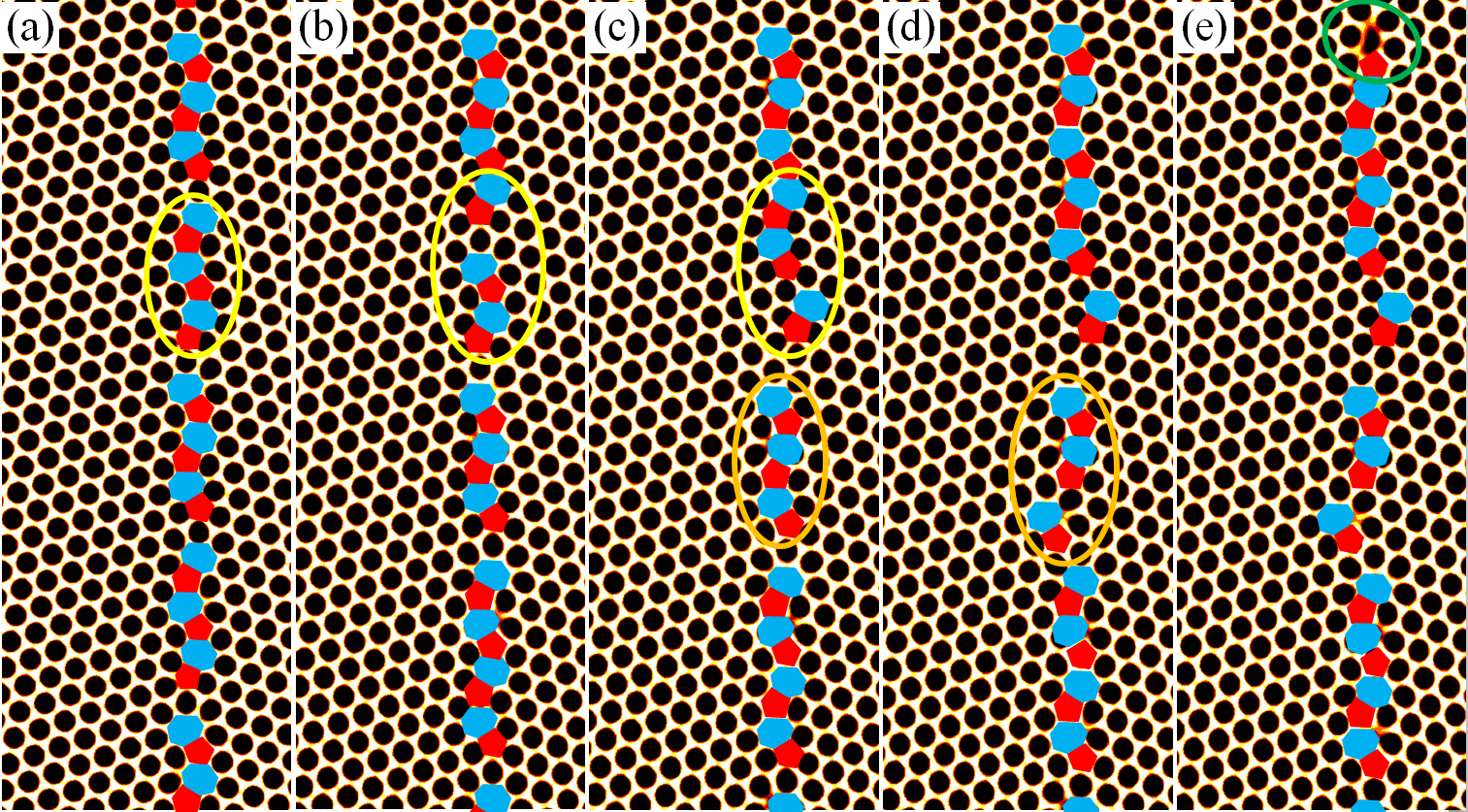}
  \includegraphics[width=0.32\textwidth]{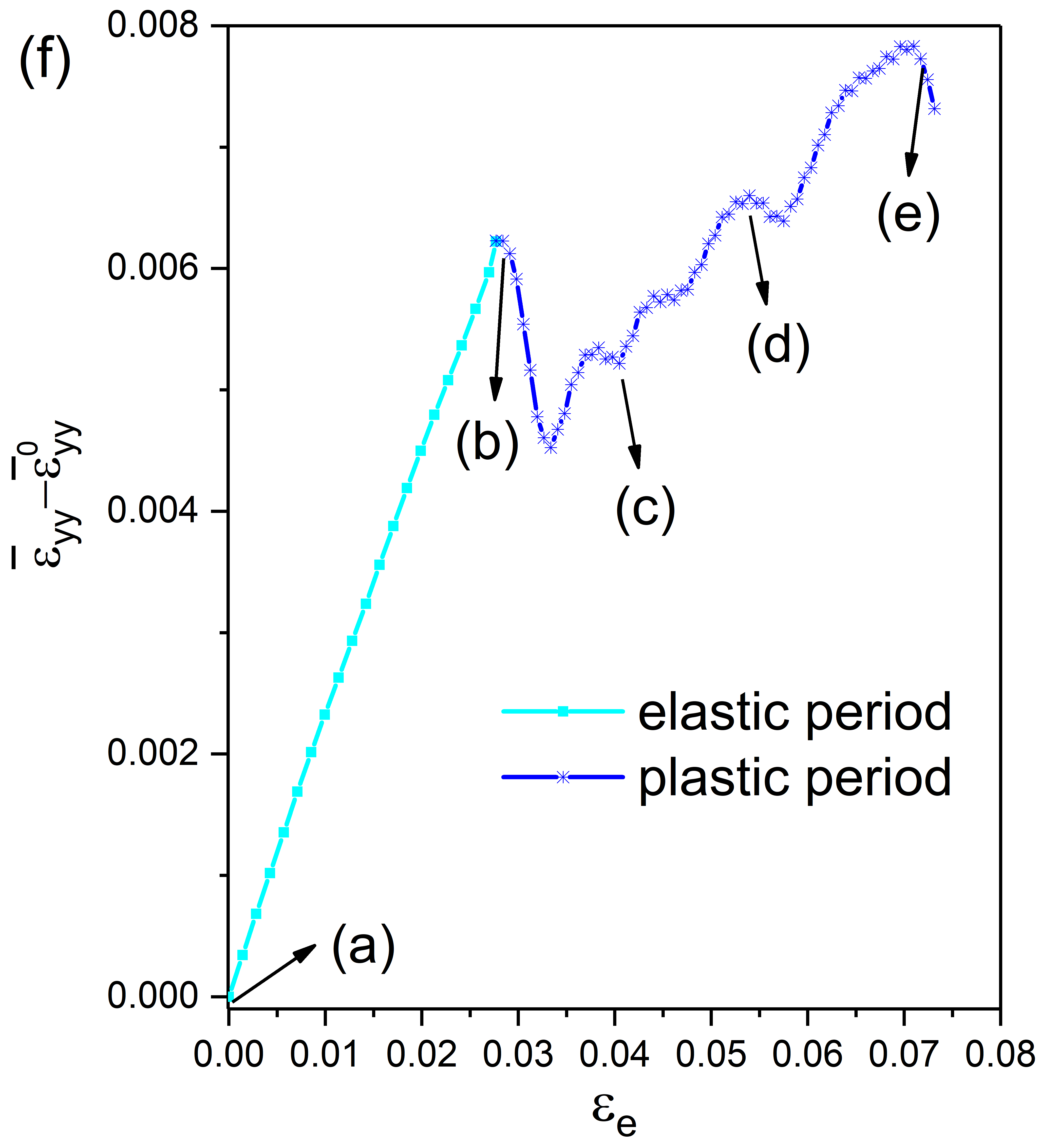}
  \caption{Snapshots of the structural evolution process of ZZ3 GB, at $\epsilon=-0.1$ ($T=4278$ K)
    and $\varepsilon_e=0$ (a), $2.91\%$ (b), $4.12\%$ (c), $5.11\%$ (d), and $7.18\%$ (e).
    Pentagons and heptagons are colored in red and blue, respectively. Only a portion of the
    simulated system is shown here. The average local strain vs. $\varepsilon_e$ is given in (f).}
  \label{fig:ZZ3_e01}
\end{figure*}

Details of sample failure and deformation dynamics at ultrahigh temperature $\epsilon=-0.1$
(with $T=4278$ K, close to the melting point of graphene which corresponds to $\epsilon=0$
and $T_m=4510$ K \cite{LosPRB15}) are presented in
Figs.~\ref{fig:deformation_e01}--\ref{fig:ZZ3_e01}, showing qualitatively different behaviors
of sample yielding as compared to those of relatively lower temperature given above in
Sec.~\ref{sec:mech_brittle}. For all the AC GBs a similar failure behavior, characterized by
the formation of nanovoids, is found at late stages, with some snapshots shown in
Fig.~\ref{fig:deformation_e01} (a)--(c). Different from the lower-temperature results,
here plastic deformation occurs after the beginning elastic regime and is characterized by
a jerky or serrated flow behavior and stick and climb-glide type dislocation motion (i.e.,
alternating periods of GB dislocation migration and temporary stoppage or pinning), with
the corresponding yield process detailed in Fig.~\ref{fig:AC1_e01} for the example of AC1
sheet. This jerky plastic flow can be identified from the result of
the local strain $\varepsilon_{yy}$ averaged within the active zone of simulation, i.e.,
$\bar{\varepsilon}_{yy}$. Its variation with respect to the initial average strain
$\bar{\varepsilon}_{yy}^0$ at $\varepsilon_e=0$, $\bar{\varepsilon}_{yy}-\bar{\varepsilon}_{yy}^0$,
is plotted against the imposed engineering strain $\varepsilon_e$ in Fig.~\ref{fig:AC1_e01}(a).
After the imposed strain reaches the yield point with strain value $\varepsilon_e^Y$ (e.g.,
$\varepsilon_e^Y = \varepsilon_e^1 = 3.37\%$ in Fig.~\ref{fig:AC1_e01}), corresponding to
the end of elasticity regime and the onset of jerky plasticity, a steep drop in the
$\bar{\varepsilon}_{yy}$ curve occurs. This drop of local strain/stress is attributed to
the glide-climb motion of all the $5|7$ dislocations at the GB simultaneously, as can be
seen by comparing the corresponding structure snapshots at the two ends of the drop
$\varepsilon_e = \varepsilon_e^1$ and $\varepsilon_e^2$ in the insets of Fig.~\ref{fig:AC1_e01}(a).
During this period each dislocation climbs down one atomic step and also glides one atomic step
to the right. After then dislocations are pinned by the underlying lattice structure (i.e.,
the effect of Peierls barriers) and become immobile, and the applied uniaxial stress/strain
accumulates in the system, with the average build-up rate similar to that of the elastic
regime (see the blue stick period in Fig.~\ref{fig:AC1_e01}(a)) given the absence of dislocation
motion. Once the pinning barriers are overcome by the accumulated stress the unpinned dislocations
move again, by further climbing down one atomic step but gliding one step back to the left (i.e.,
the opposite direction of glide compared to the previous period of motion; see the $\varepsilon_e^3$
inset). This causes the second steep drop of the average local strain. Compared this state (at
$\varepsilon_e = \varepsilon_e^3$) to the onset of plastic deformation (at $\varepsilon_e =
\varepsilon_e^1$), the net motion of dislocation is the climb of two atomic steps down, but no
net glide. This procedure of stick-and-climb-glide is repeated until the GB failure occurs,
showing as the nanovoid formation at some GB dislocations (see $\varepsilon_e^4$ vs.
$\varepsilon_e^5$ in Fig.~\ref{fig:AC1_e01}(a)). The corresponding spatial distribution of
the local strain $\varepsilon_{yy}$ is given in Fig.~\ref{fig:AC1_e01}(b) and (c), at the
onset of jerky plasticity and the formation of nanovoid respectively. Note that this behavior
of dislocation migration only appears at high enough temperature, with fast enough atomic
diffusion process and low enough defect pinning barriers such that the dislocation motion
would occur at early enough stage before sample failure (i.e., before nanovoids form at
the GB).

The situation for ZZ GBs at this ultrahigh temperature is more complicated, with the deformation
processes depending on the GB misorientation angle and details of the dislocation structure (see
Fig.~\ref{fig:deformation_e01} (d)--(f) for some snapshots at large strains with $\epsilon=-0.1$
or $T=4278$ K). For low and intermediate angles (ZZ1 and ZZ2 GBs with $\theta=8.61^\circ$ and
$13.17^\circ$) a similar behavior is found, which however is different from the high-angle one
(ZZ3 with $\theta=23.49^\circ$). At large enough strains (beyond the yield point), in both ZZ1
and ZZ2 sheets the motion of dislocations under the tensile load causes the split of the
dislocation array at each GB along the direction perpendicular to the external loading,
leading to two sub-arrays of dislocations of $(1,0)$ and $(0,1)$ types respectively. This
is analogous to the phenomenon of collective dislocation separation from the GBs (see
Fig.~\ref{fig:deformation_e01} (d) and (e)), but different from the GB disintegration at
lower-temperature ZZ1 GBs (Fig.~\ref{fig:fracture_e05}(d) with $\epsilon=-0.5$ or $T=3352$ K)
which instead occurs after GB failure and is accompanied by the motion of dislocation-bound
nanovoids.

Details of the deformation dynamics for ultrahigh-temperature ZZ1 GBs are given in
Fig.~\ref{fig:ZZ1_e01}, showing the behavior of jerky plastic flow similar to that of AC sheets,
as seen from the corresponding strain curve of $\bar{\varepsilon}_{yy}$ vs. $\varepsilon_e$.
However, in the plastic regime the dynamics of mobile dislocations is different. As shown in
Fig.~\ref{fig:ZZ1_e01}(a), during the first drop of local strain/stress from the yield point
$\varepsilon_e = \varepsilon_e^1$ to $\varepsilon_e^2$ the $(1,0)$ dislocations migrate via a
mixed motion of glide and climb (one atomic step leftward and one step upward for the example of
Fig.~\ref{fig:ZZ1_e01}(a)), while the $(0,1)$ dislocations remain immobile. After the subsequent
stick period with the pinning of both types of defects, the next drop in strain curve occurs,
which is caused by the glide and climb of $(0,1)$ dislocations (one atomic step rightward and one
step upward if comparing $\varepsilon_e^3$ and $\varepsilon_e^2$ inset panels), while the $(1,0)$
dislocations keep immobile this time. The repeat of this procedure, as featured by the alternative
motion of one type of dislocations intermitted by a temporary period of defect pinning, results
in the splitting of each GB dislocation array into two sub-arrays as described above and also
illustrated in Fig.~\ref{fig:ZZ1_e01} (b) vs. (c) in terms of local strain distribution. Compared
to the case of AC GBs shown in Fig.~\ref{fig:AC1_e01}, here the jerky behavior of plastic flow
is less regular, with shorter stick/pinning periods as time (or $\varepsilon_e$) increases.
This can be attributed to larger degree of dislocation glides for this type of ZZ GBs (given
its different dislocation structure and GB strain distribution compared to the AC GBs; see
Fig.~\ref{fig:GBs}) and the interaction between dislocations separated from two neighboring GBs.
At late times (with large applied $\varepsilon_e$) the dislocation sub-arrays from two different
GBs will annihilate, leading to a fast decrease of system elastic energy and stress.

Among the ZZ sheets the detailed dynamics of dislocations highly depends on the type of
GB defect structure. As shown in Fig.~\ref{fig:GBs}(f) and also Fig.~\ref{fig:ZZ3_e01}(a),
the high-angle ZZ3 GB is composed of a different type of dislocation array (i.e., connected
dislocation triplets) and exhibits different strain distribution as compared to the
low/intermediate-angle ZZ1 and ZZ2 GBs with disperse $(1,0)$ and $(0,1)$ $5|7$ defects,
leading to more complex deformation behavior under the imposed tension. An example of a
series of dislocation dynamics under tensile stress is illustrated in Fig.~\ref{fig:ZZ3_e01}
(a)--(e). Beyond the yield point a $(0,1)+(1,0)+(0,1)$ dislocation triplet breaks up within
this portion of the GB shown in the figure, and its top $(0,1)$ dislocation glides and climbs
up one atomic step to join the neighboring dislocation cluster (see the circled region in
Fig.~\ref{fig:ZZ3_e01} (a) and (b)). Once the applied stress is built up and exceeds the
pinning barrier, the remaining pair breaks up again, where the $(1,0)$ dislocation migrates
one step to connect with the upper cluster (see the upper circle in Fig.~\ref{fig:ZZ3_e01}(c)).
Further increasing the stress leads to the disintegration of another dislocation triplet, i.e.,
the $(1,0)+(0,1)+(1,0)$ one circled in the lower part of Fig.~\ref{fig:ZZ3_e01} (c) and (d).
At high enough external strain/stress nanovoids form at the GBs, as indicated in
Fig.~\ref{fig:ZZ3_e01}(e), corresponding to the occurrence of sample failure. Similar
procedure of dislocation triplet break-up and single dislocation migration occurs in various
segments of the GB (see also Fig.~\ref{fig:deformation_e01}(f)), and the corresponding plot
of strain curve showing a jerky plastic flow is given in Fig.~\ref{fig:ZZ3_e01}(f).

A similar behavior of jerky plasticity has been obtained in previous
experimental and MD studies of GBs in 3D nanowires or micropillars under uniaxial compression
\cite{TuckerMSMSE13,ImrichActaMater14}, although with different underlying mechanisms.
In those 3D cases new dislocations nucleate inside the system under large enough built-up stress
and then migrate, while the stick period (without any dislocation activities) is caused by
either dislocation starvation (which corresponds to larger amplitude of jerky behavior) or
pinning \cite{TuckerMSMSE13}. In the 2D system studied here, no nucleation of additional
dislocations is found, and all the dislocations are those from the initial state; the
phenomenon of high-temperature jerky plastic flow originates purely from stress-induced
dislocation motion and temporary lattice pinning.

\subsection{The brittle-to-ductile transition}
\label{sec:transition}

\begin{figure*}
  \centering
  \includegraphics[width=0.60\textwidth]{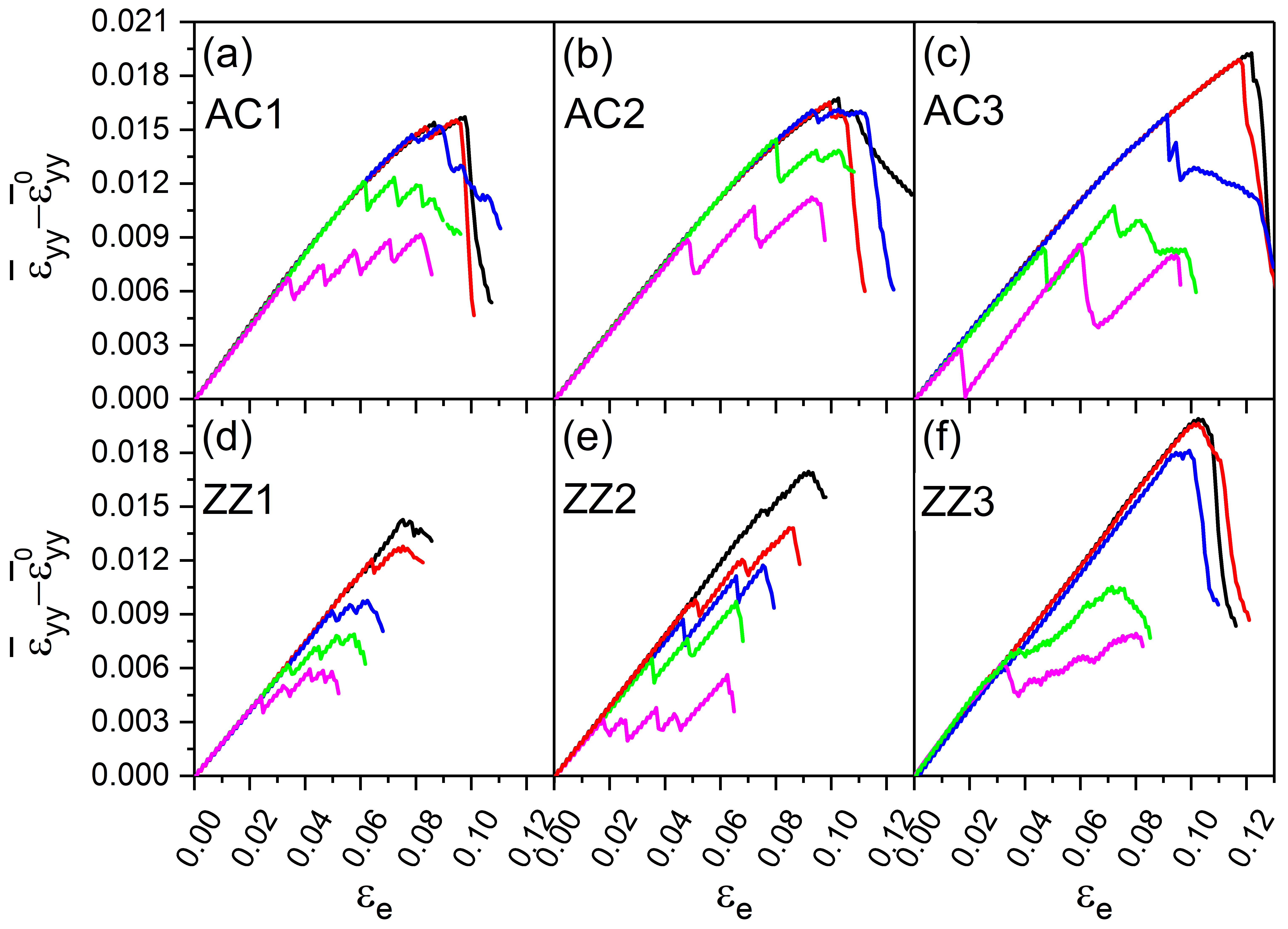}
  \includegraphics[width=0.33\textwidth]{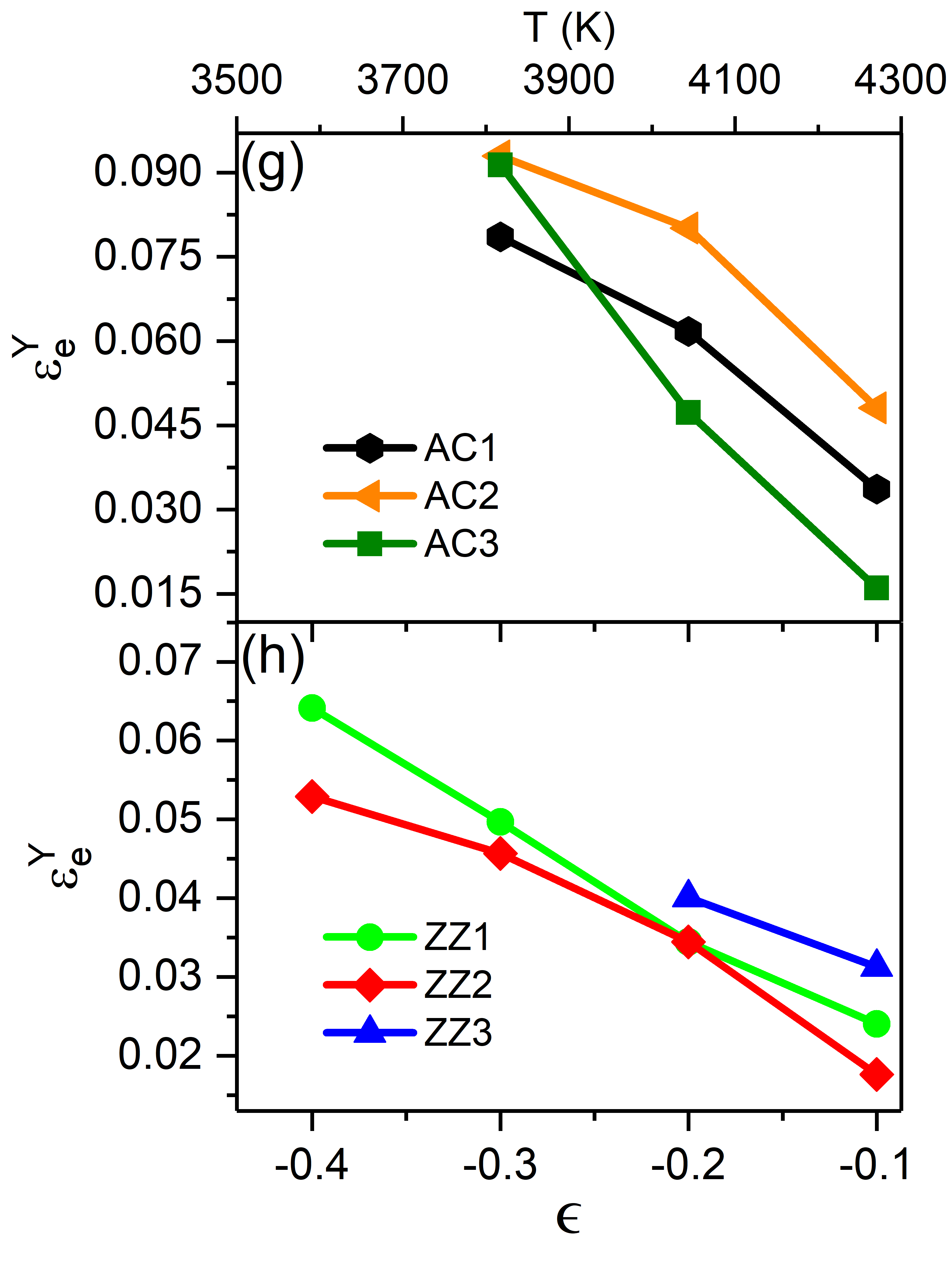}
  \caption{(a)--(f): Change of average local strain, $\bar{\varepsilon}_{yy}-\bar{\varepsilon}_{yy}^0$,
    as a function of $\varepsilon_e$ for AC and ZZ sheets with the direction of tensile loading
    parallel to the GBs, at different values of temperature $\epsilon=-0.5$ ($T=3352$ K;
    black), $-0.4$ ($3584$ K; red), $-0.3$ ($3815$ K; blue), $-0.2$ ($4047$ K; green), and
    $-0.1$ ($4278$ K; magenta) from top to bottom. (g) and (h): Values of $\varepsilon_e$ at
    the yield point, $\varepsilon_e^Y$, as a function of temperature, for AC and ZZ GBs respectively.}
  \label{fig:strain_e01-05}
\end{figure*}

\begin{figure*}
  \centering
  \includegraphics[width=0.7\textwidth]{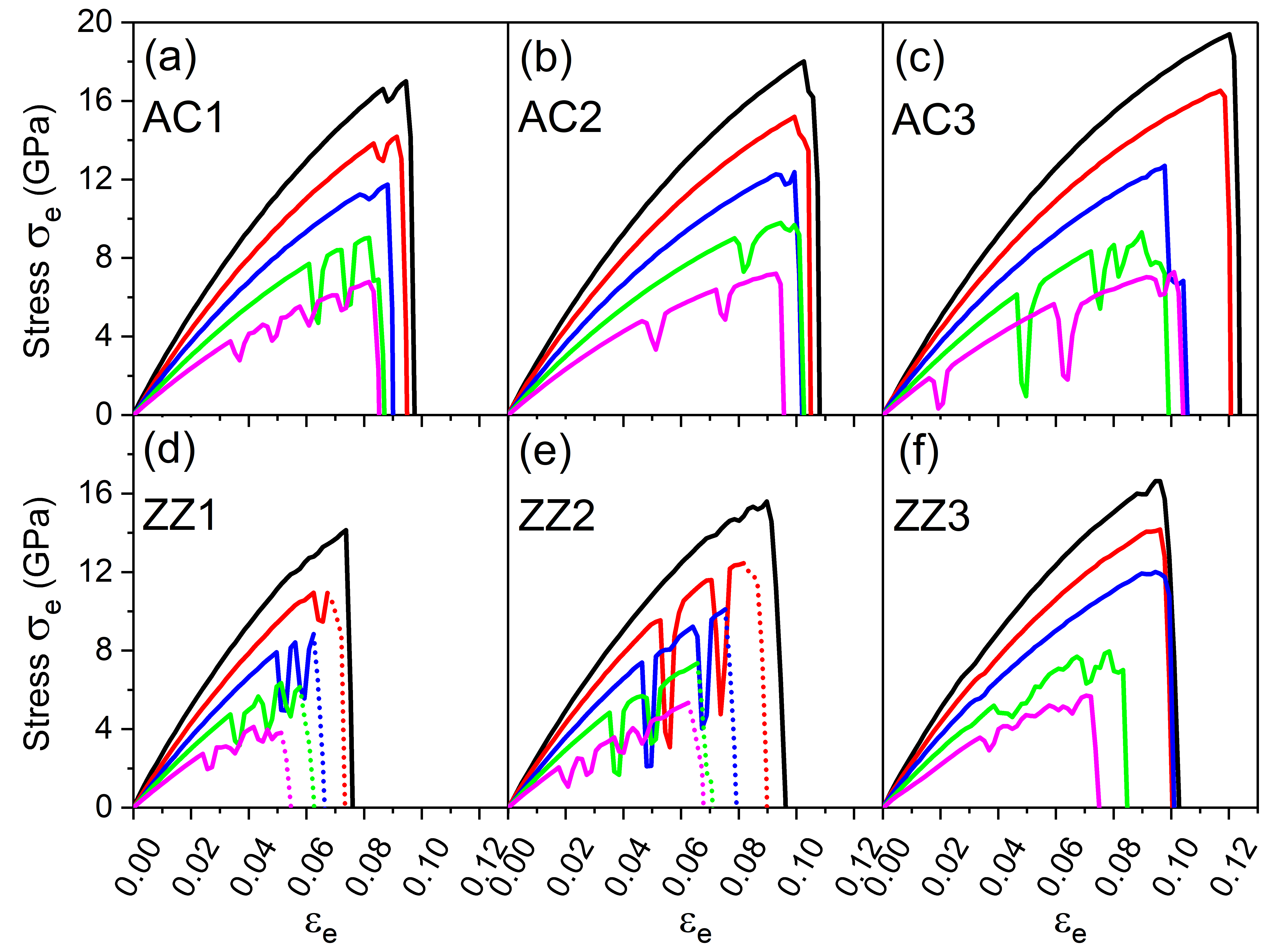}
  \caption{Stress-strain curves for AC ((a)--(c)) and ZZ ((d)--(f)) sheets with the direction
    of tensile loading parallel to the GBs, at different values of temperature
    $\epsilon=-0.5$ ($T=3352$ K; black), $-0.4$ ($3584$ K; red), $-0.3$ ($3815$ K; blue),
    $-0.2$ ($4047$ K; green), and $-0.1$ ($4278$ K; magenta) from top to bottom. The dotted
    lines in (d) and (e) correspond to fast decrease of system elastic energy and stress caused
    by the annihilation of dislocation sub-arrays separated from two different GBs given the
    finite system size in the simulation.}
  \label{fig:stress_e01-05}
\end{figure*}

To summarize the effect of system temperature on the GB deformation dynamics, we
plot the strain-strain curves of $\bar{\varepsilon}_{yy}$ vs. $\varepsilon_e$ in
Fig.~\ref{fig:strain_e01-05} for all the GBs examined here with temperature ranging
from $\epsilon=-0.5$ ($T=3352$ K) to $-0.1$ ($4278$ K). The corresponding stress-strain
curves are given in Fig.~\ref{fig:stress_e01-05}. At low enough temperatures
($T \lessapprox 3352$ K) the brittle-type behavior of fracture is observed, analogous
to previous results of MD simulations. As temperature increases a transition to plastic
or ductile behavior occurs, particularly the behavior of jerky plastic flow at high
enough temperatures (e.g., $T=4278$ K studied above). This phenomenon of jerky plasticity
can be seen from the serrated or seemingly irregular curves of the stress-strain relation
given in Fig.~\ref{fig:stress_e01-05}. Each turning point in the stress-strain curve can be
matched to the corresponding point of the strain-strain curve in Fig.~\ref{fig:strain_e01-05},
which in turn has been one-to-one correlated to the associated stick and climb-glide motion
of GB dislocations as described in detail in Sec.~\ref {sec:mech_highT} for both AC and ZZ
GBs (see also Figs.~\ref{fig:AC1_e01}--\ref{fig:ZZ3_e01}). 

Usually the brittle-to-ductile transition is accompanied by the generation of
sufficient amount of new dislocations from existing sources inside the sample. However,
in the 2D system studied here no such dislocation generation or nucleation is found;
instead, the plastic deformation, particularly jerky plasticity, develops as long as the
GB dislocations are able to migrate under large enough external stress/strain, before
the atomic-ring breaking and the nucleation of nanovoids would have time to initiate
at the GB. The motion of dislocations (both glide and climb) can occur even when
being limited around the GB region, as seen in AC and ZZ3 GBs (Figs.~\ref{fig:AC1_e01}
and \ref{fig:ZZ3_e01}), under the condition that the individual GB dislocations or
dislocation clusters (e.g., pairs or triplets) are separated from each other and the
temperature is high enough such that pinning of the lattice and binding of nearby
dislocations via elastic interaction could be overcome by large enough atomic mobility
under external stress. This can be made possible in graphene-type 2D systems which,
different from  more complicated 3D cases, do not involve dislocation loops and
intersection and are hence less restricted in terms of dislocation motion.

Details of this brittle to ductile/plastic transition would then depend on the specific
structure of GBs and dislocations, and thus the type of GB (AC vs. ZZ) and the tilt
angle, which correspond to different strain distribution and binding barriers. A similar
temperature range of transition is found for all three AC GBs examined, for which the
plastic deformation starts to develop at late stage (i.e., large applied strain
$\varepsilon_e$) at $\epsilon=-0.3$ ($T=3815$ K), as can be seen more clearly from the
strain curves in Fig.~\ref{fig:strain_e01-05} (a)--(c). This is attributed to their similar
dislocation structure at the GBs (i.e., arrays of disperse $(1,0)$ dislocations as shown
in Fig.~\ref{fig:GBs}). On the other hand, at ultrahigh temperatures the stick period of the
jerky plastic flow (i.e., the waiting time) is longer for larger tilt angle, as a result
of higher dislocation density at the GB and thus stronger barrier for dislocation motion.
For ZZ GBs shown in Figs.~\ref{fig:strain_e01-05} and \ref{fig:stress_e01-05} (d)--(f),
the temperature range for brittle-plastic transition is different from that of AC cases
and varies with the GB angle, given different GB dislocation structure and arrangement
particularly between ZZ3 and ZZ1 or ZZ2 (dislocation triplet vs. disperse ones). Jerky
plasticity occurs at much lower temperature for ZZ1 and ZZ2 with low and intermediate tilt
angles ($\epsilon \gtrapprox -0.4$ or $T \gtrapprox 3584$ K), showing larger amplitude of
flow serration, as compared to that of ZZ3 with high angle (at $\epsilon \simeq -0.2$ or
$T \simeq 4047$ K). This is related to denser dislocation arrays in ZZ3 and more complex
process of dislocation triplet break-up and migration (see Fig.~\ref{fig:ZZ3_e01}), giving
a smoother plastic flow that needs higher temperature to initiate. For both AC and ZZ GBs
the onset of jerky plastic flow (or the yield point) appears earlier at smaller strains
for higher temperature, as expected, with results shown in Fig.~\ref{fig:strain_e01-05} (g)
and (h) for the yield-point values of $\varepsilon_e^Y$ as a function of temperature.

\subsection{Effects of strain rate and pulling direction}

\begin{figure*}
  \centering
  \includegraphics[width=0.49\textwidth]{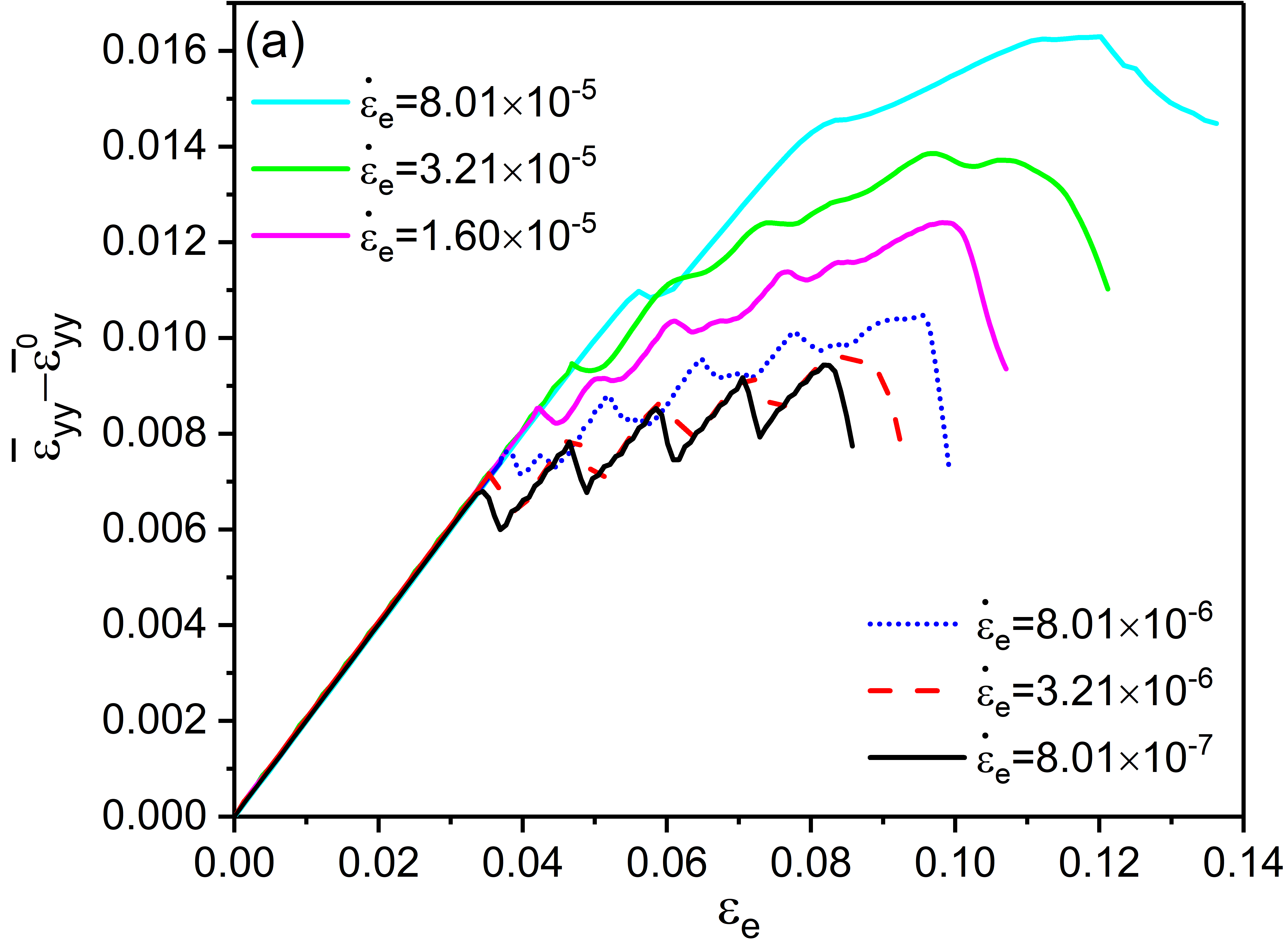}
  \includegraphics[width=0.45\textwidth]{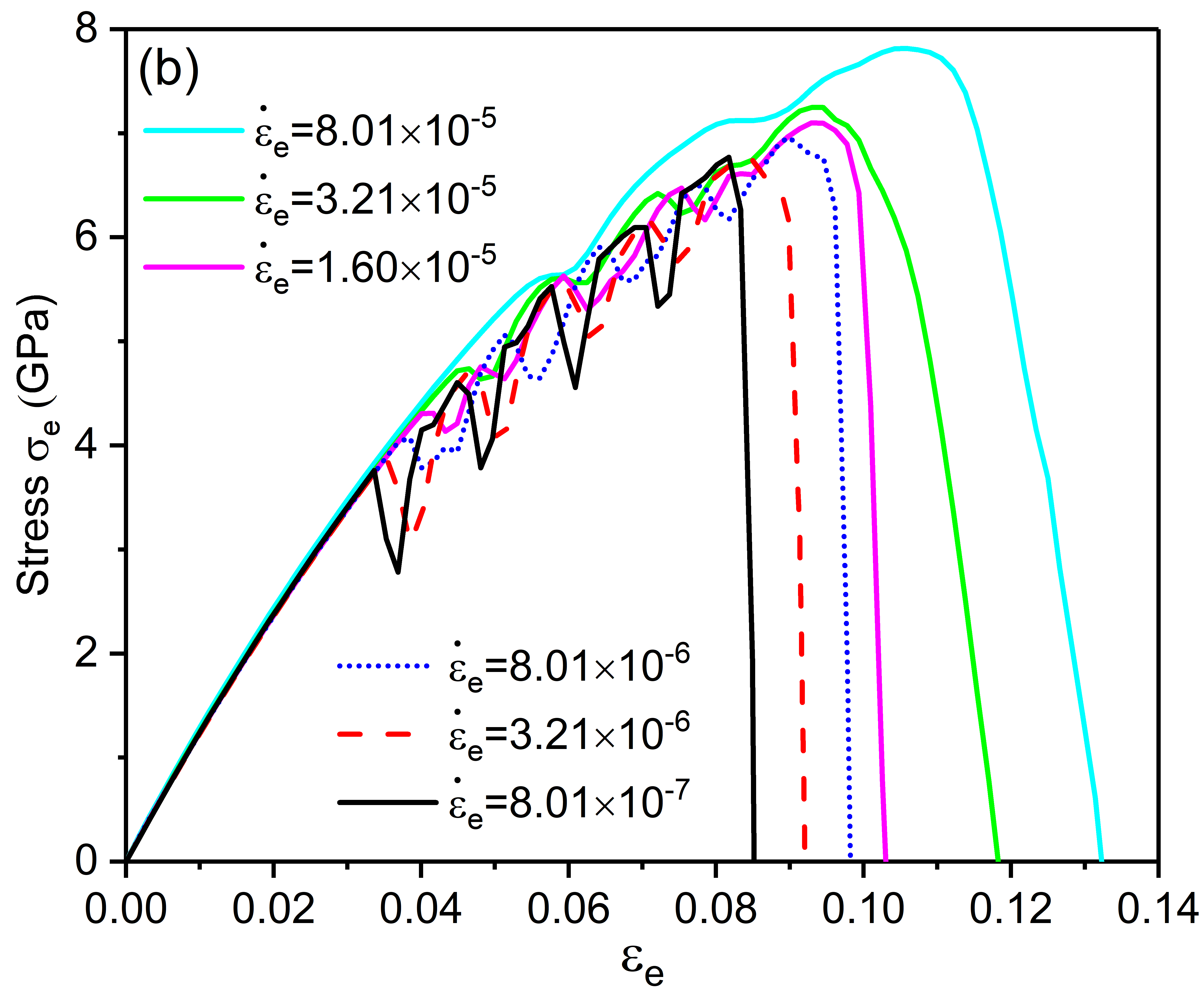}
  \caption{Effect of the applied strain rate $\dot{\varepsilon}_e$ on (a) the change of
    average local strain $\bar{\varepsilon}_{yy}-\bar{\varepsilon}_{yy}^0$ vs. applied strain
    $\varepsilon_e$ and (b) the stress-strain relation, for AC1 sheet at $\epsilon=-0.1$
    ($T=4278$ K) that is subjected to tensile loads parallel to the GB.}
  \label{fig:strain_rates}
\end{figure*}

In the calculations given above, a small strain rate of tensile deformation,
$\dot{\varepsilon}_e = 8.01 \times 10^{-7}$, is applied. The choice of small enough
strain rate is important for achieving the behavior of jerky plasticity, as evidenced
in Fig.~\ref{fig:strain_rates} which show the results of strain-strain and stress-strain
curves when the strain rate $\dot{\varepsilon}_e$ is varied over two orders of magnitudes,
i.e., spanning from $8.01 \times 10^{-5}$ to $8.01 \times 10^{-7}$ for the example of AC1
sheet at $\epsilon=-0.1$ with $T=4278$ K. The jerky behavior becomes apparent when
$\dot{\varepsilon}_e \lessapprox 8.01 \times 10^{-6}$ for this example, and too large
strain rate would lead to brittle-type failure with limited or even the absence of plastic
flow, as accompanied by very little degree of dislocation motion (as observed in our PFC
simulation at e.g., $\dot{\varepsilon}_e = 8.01 \times 10^{-5}$). It can be understood from
the control mechanism of jerky plastic flow described above in Sec. \ref{sec:transition}
in terms of GB dislocation dynamics. Between two subsequent pulling steps a sufficient
evolution time (on the order of atomic diffusion timescales) is needed to allow the completion
of dislocation migration process, leading to the requirement of small enough strain rates.
The simulation method then needs to simultaneously incorporate the fast-timescale elastic
relaxation and slow-scale atomic or vacancy diffusion process, a condition that is
satisfied by the PFC modeling adopted here (which combines the IPFC scheme for fast
mechanical relaxation and the diffusive dynamics as seen from Eq.~(\ref{eq:timescale})
in \ref{sec:parameterization} for standard PFC timescale that is inversely proportional to
vacancy diffusion constant). This can also explain the difficulty of identifying plastic
deformation in MD simulations for graphene: The standard atomistic technique like MD is
of atomic vibration timescales, and the strain rate used in MD simulations of graphene
is on the order of $10^{-4}$--$10^{-2}$ ps$^{-1}$ \cite{Grantab946,Wei12,YI2013373}.
This fast strain rate, which needs to be used in atomistic simulations with reasonable
computational costs, is far from reaching the requirement of diffusive timescales at
each tensile stretching step to facilitate the climb-glide motion of dislocations,
particularly climb and the dislocation migration along the direction different from
(e.g., perpendicular to) the external loading that is crucial for the occurrence of
jerky plastic and ductile behavior at high enough temperature.

\begin{figure*}
  \centering
  \includegraphics[width=0.36\textwidth]{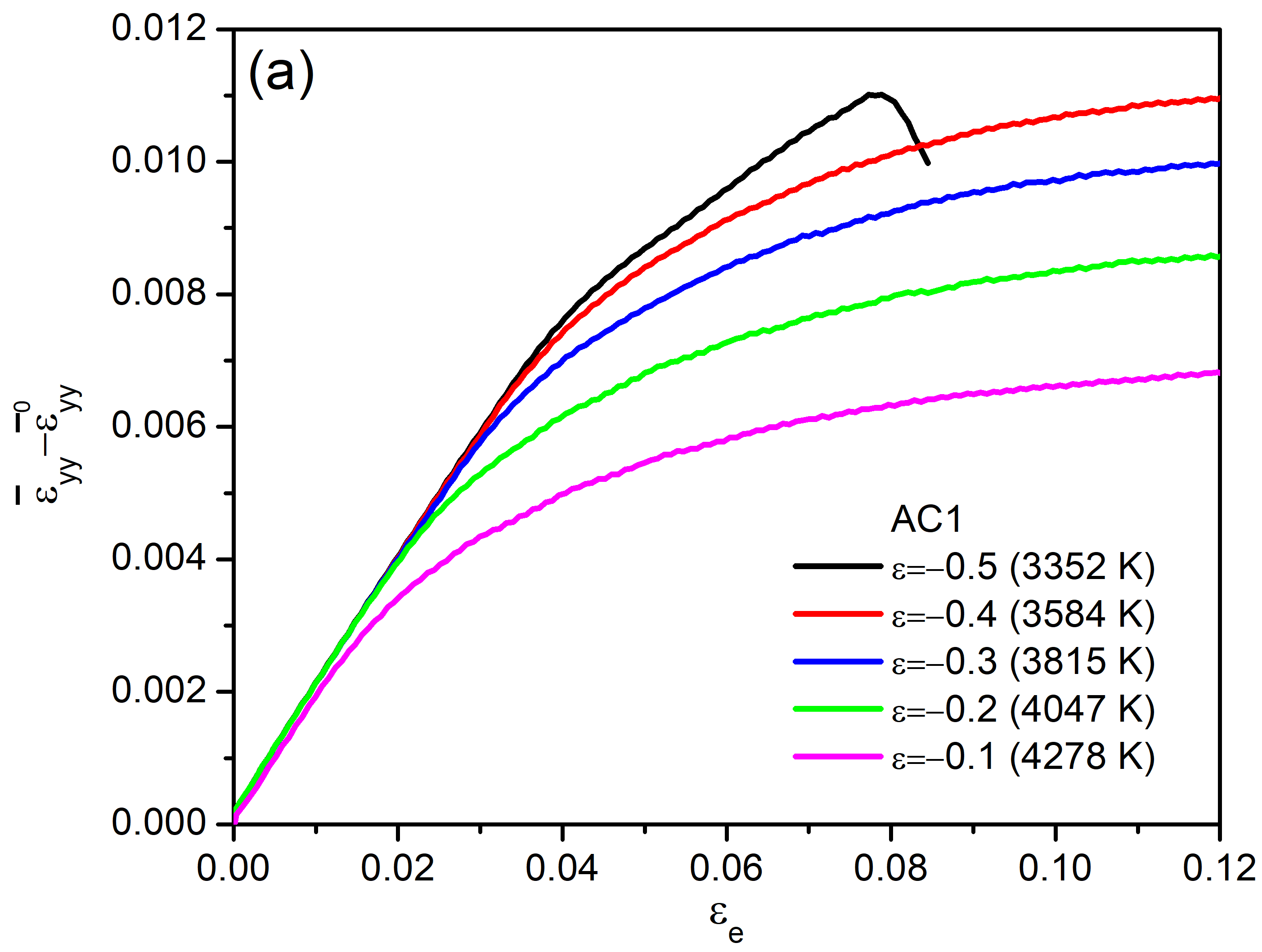}
  \includegraphics[width=0.33\textwidth]{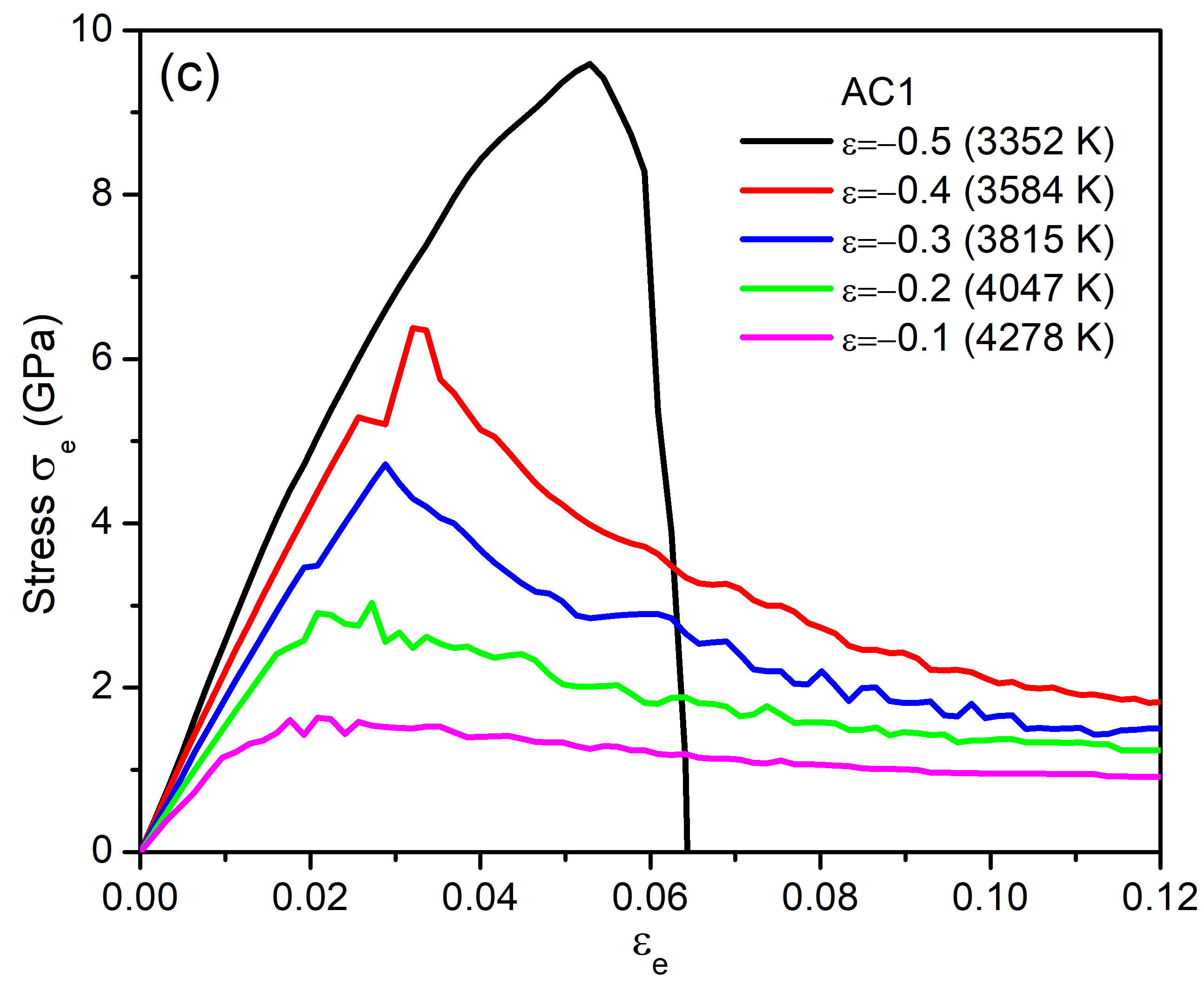} \\
  \includegraphics[width=0.36\textwidth]{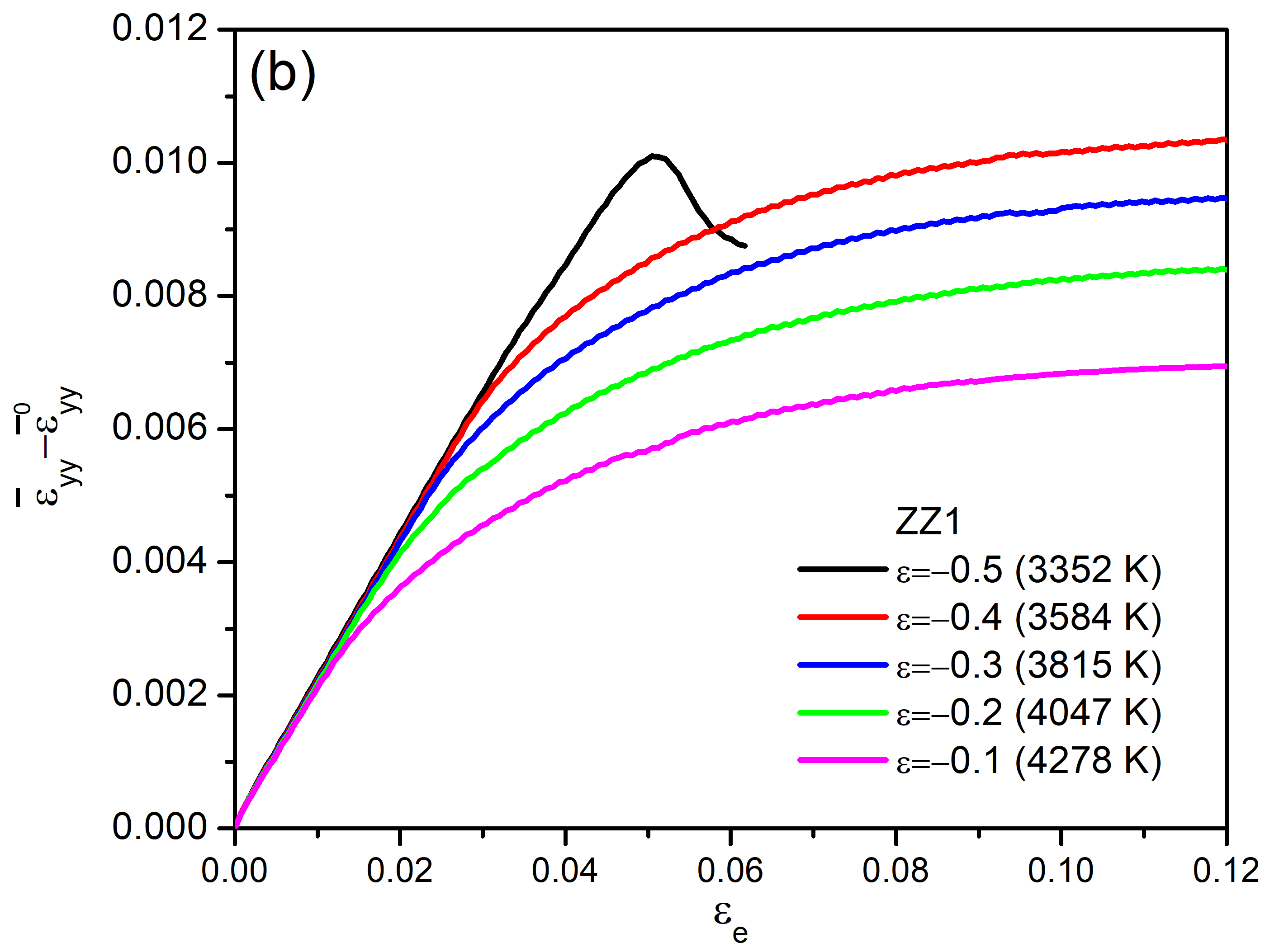}
  \includegraphics[width=0.33\textwidth]{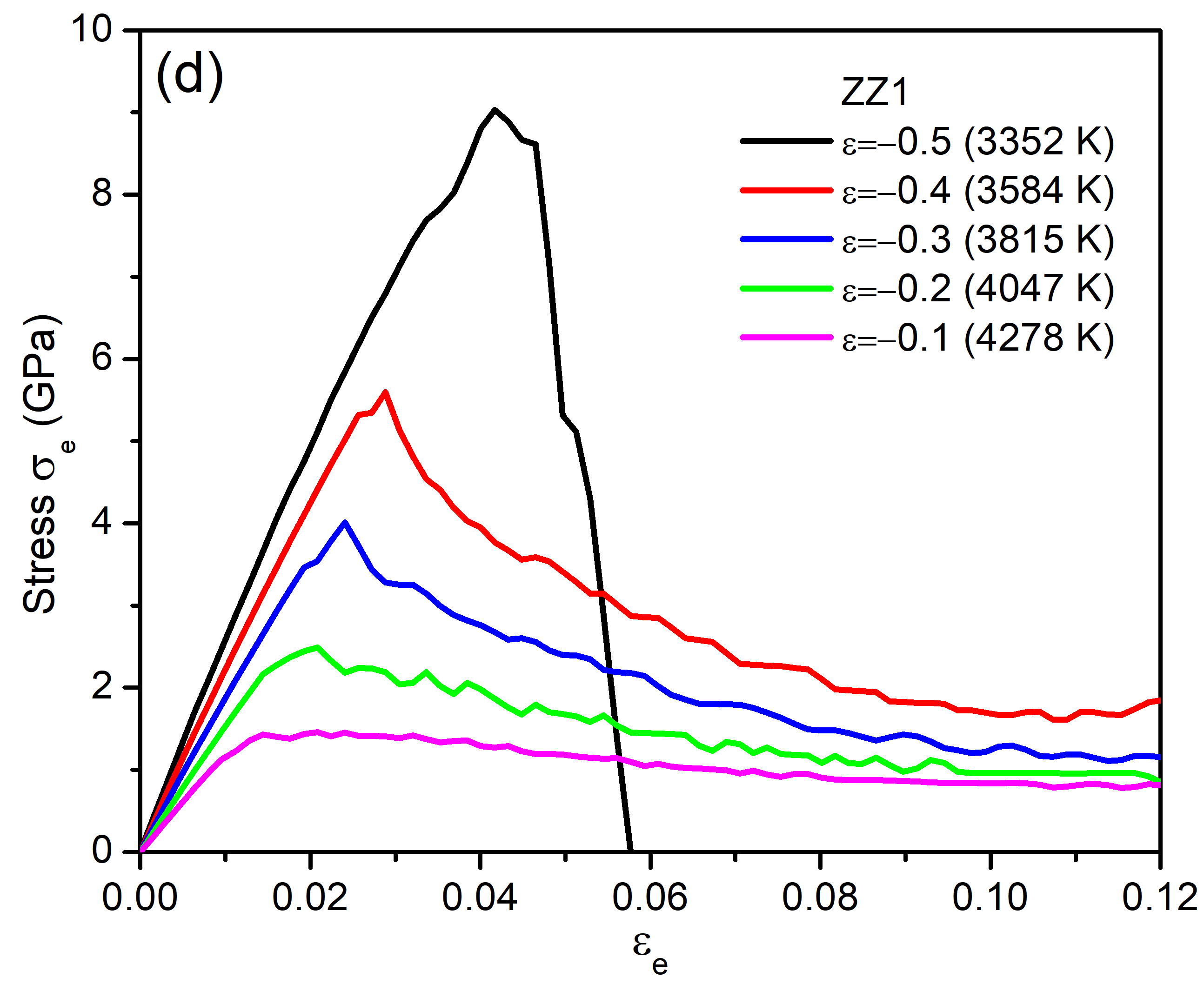}
  \caption{Change of average local strain, $\bar{\varepsilon}_{yy}-\bar{\varepsilon}_{yy}^0$,
    as a function of $\varepsilon_e$ for (a) AC1 and (b) ZZ1 sheets, with the direction of
    tensile loading perpendicular to the GBs under the strain rate $\dot{\varepsilon}_e =
    8.01 \times 10^{-7}$ at different temperatures. The corresponding stress-strain curves
    are given in (c) and (d), respectively.}
  \label{fig:perpendicular}
\end{figure*}

It is noted that all the above results are for the tensile loading applied parallel
to the GB line. To examine the effect of external pulling direction, we conduct
additional simulations for tensile loads perpendicular to the GB. The PFC system
setup is similar to that of Fig.~\ref{fig:GB_structure}, other than a single GB
located at $y=L_y/2$ in the middle of the solid region separating two misoriented
grains. The GB line is aligned along the $x$ direction, perpendicular to the
vertical ($y$) direction of pulling. Some sample results for low-angle GBs (AC1
and ZZ1) are given in Fig.~\ref{fig:perpendicular}, across the same temperature range
as above. As seen from both the average local strain variation and the stress-strain
curves, similar to the above case of parallel pulling a behavior of brittle fracture
is obtained at temperature $\epsilon=-0.5$ ($T=3352$ K), while a brittle-to-ductile
transition is observed at higher temperature (starting at $\epsilon=-0.4$ with $T=3584$ K
in Fig.~\ref{fig:perpendicular}). However, here a different property of plastic flow is
found, with much smoother and more irregular behavior compared to that shown in
Figs.~\ref{fig:strain_e01-05} and \ref{fig:stress_e01-05} for the parallel case.
This can be attributed to different behavior of high-temperature dislocation motion
before the sample failure, which is observed in our simulations to be restricted only
along the GB line when the system is stretched perpendicular to the GB. In real
polycrystalline samples with multiple grain orientations, any given pulling direction
of the whole sample will correspond to different loading directions for different GBs,
including both the parallel and perpendicular ones and mostly the scenarios in between,
such that the system mechanical behavior will be the combined effect of them. In the
majority of cases the GBs would not be oriented perpendicular (or nearly perpendicular)
to the pulling direction, and thus the dislocation motion would not be constrained along
the GB line, a scenario that is relatively closer to the case studied above with
GB-parallel loading.

\section{Conclusions}

We have applied the phase field crystal method, which is combined with an interpolated
scheme that enables fast mechanical relaxation, to examine the mechanical response of
2D graphene bicrystals across the high or ultrahigh temperature range that is beyond
most of previous experiments and atomistic simulations. Small enough strain rate is
used in our simulations, which is important for accessing the plastic deformation regime
that is missing in atomistic studies of graphene. The systems investigated include both
armchair and zigzag types of symmetric tilt grain boundaries that are subjected to
uniaxial tensile loads, for low, intermediate, and high misorientation angles. For the GBs
studied here, although the associated equilibrium dislocation structures and local strain
distribution are mostly independent of temperature change in the absence of external stress,
their mechanical behavior and deformation dynamics vary significantly when reaching high
enough temperature. Even in a relatively high temperature range (around $3350$ K), results
of mechanical deformation and brittle fracture that are similar to those of previous
low-temperature MD simulations and experiments can be obtained, including nanovoid
formation and crack initiation at the locations of GB dislocations and the subsequent
crack propagation. Our PFC simulations also reveal an additional type of failure dynamics
for low-angle zigzag GBs, showing as the disintegration and splitting of GB dislocation
array through the migration of nanovoid-bound dislocations. An increase of temperature
leads to the decrease of ultimate tensile strength and Young's modulus for both AC and
ZZ GB sheets, and more importantly, results in a transition to ductile and plastic
deformation under tensile loads either parallel or perpendicular to the GB. The behavior
under GB-parallel loading at high enough temperature (ranging from $3584$ K to $4278$ K
for different types of GBs studied here) is characterized by jerky plasticity which is
absent in all the previous studies of graphene. This jerky plastic or stick-climb-glide
type behavior, as seen in both the average local strain curves and the stress-strain
relation, is caused by the motion of GB dislocations (in the form of either glide-mediated
climb for AC GBs or mixed motion of glide and climb for ZZ GBs) interrupted by periods of
pinned and immobile defects.

The occurrence of this temperature-induced transition between brittle fracture and
jerky plastic behavior depends on the type of grain boundary (armchair vs. zigzag) and
the misorientation angle, particularly the type of dislocation structure and arrangement
at the boundary and its local strain distribution. Microstructurally this is related to
the competition between the stress-induced formation of nanovoids at the GB and the ability
or degree of dislocation migration. The corresponding mechanisms are different from that
of traditional brittle-to-ductile transition or the phenomenon of jerky plasticity in 3D
compressed nanowires/pillars, both of which involve new dislocation nucleation that is not
found here. Instead, here the mechanisms governing the jerky plastic flow are intrinsic to
the 2D systems, where the migration of dislocations (locally around the GB) can be made
possible by the built-up stress at high enough temperature, to overcome the constraints of
pinning barriers by the lattice and the interaction of neighboring defects. Although this
study is restricted to simplified systems of bicrystals and symmetric grain boundaries,
the deformation mechanisms identified are generic and can be extended to the study of
more complex scenarios of defected or polycrystalline graphene, and thus provide further
insights into the mechanical behavior of graphene-type 2D materials particularly their
mechanical performance at high temperature.

\section*{Acknowledgments}

  Z.-F.H. acknowledges support from the National Science Foundation under Grant
  No. DMR-1609625. J.W. acknowledges support from the National Natural Science
  Foundation of China (Grants No. 51571165, 51871183), and Special Program for Applied
  Research on Super Computation of the NSFC-Guangdong Joint Fund (the second phase)
  under Grant No. U1501501. We also thank the Center for High Performance Computing
  of Northwestern Polytechnical University, China for computer time and facilities.

\appendix

\section{Parameterization of PFC model for graphene}
\label{sec:parameterization}

To obtain quantitative results of PFC simulation for real material systems, the PFC
model described in Sec. \ref{sec:model} needs to be parameterized to match graphene.
In principle the model parameterization can be conducted through the liquid-state
interparticle direct correlation functions; however, the corresponding data is usually
not available for most materials (except for a very limited number of material systems).
Extra effort is needed to quantify and convert the parameter $\epsilon$ to real
temperature $T$ and identify the corresponding conversion function $\epsilon(T)$,
which would require the calculation and fitting of two-point direct correlation at
various temperatures above but close to the melting point.

\begin{figure}
  \centerline{\includegraphics[width=0.55\textwidth]{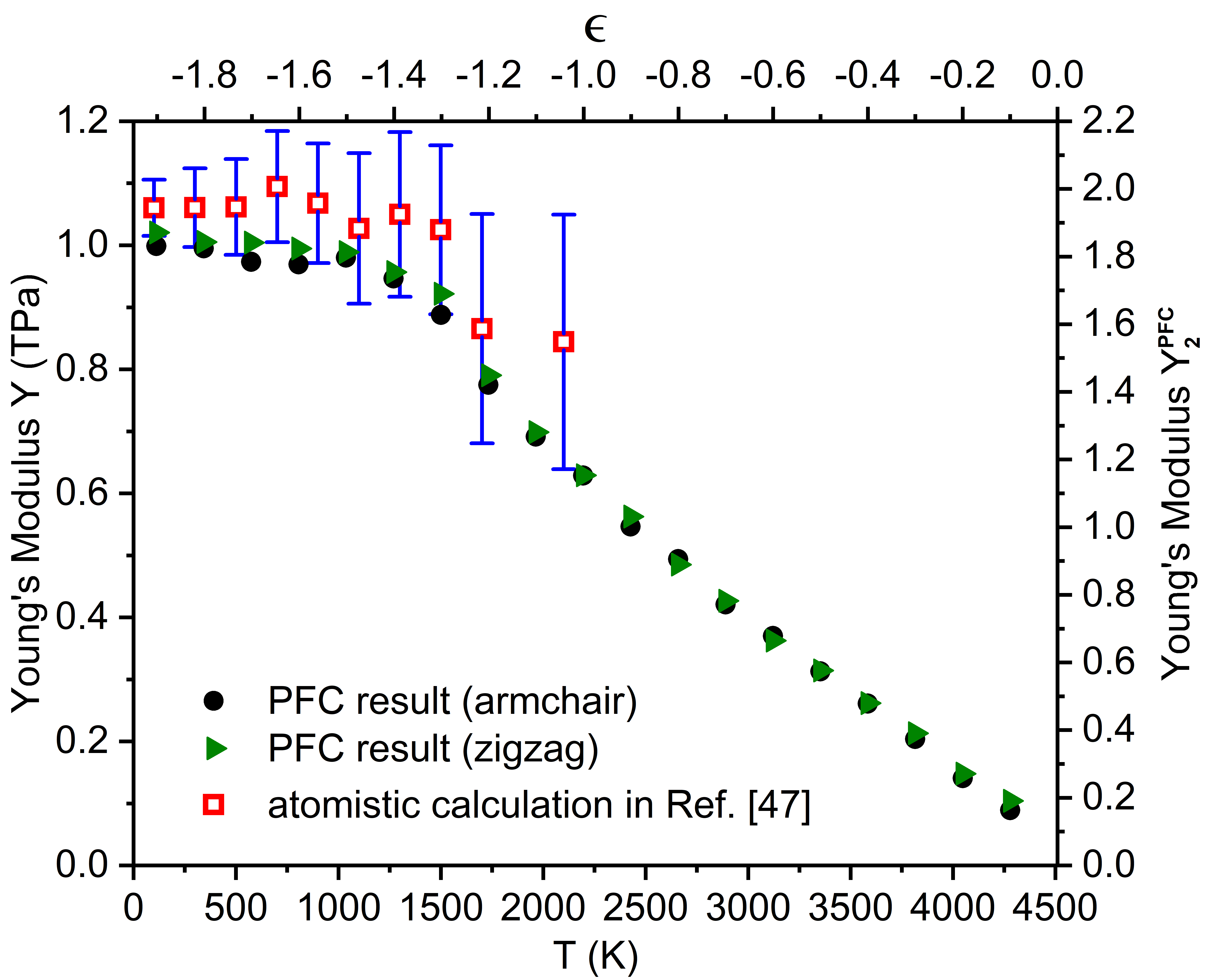}}
  \caption{Young's modulus as a function of temperature, as calculated from the PFC
    model for pristine graphene with uniaxial tensile loads along the armchair or
    zigzag direction. Both the PFC values and the corresponding real units are indicated.
    The data points obtained from atomistic calculations in Ref.~\cite{ZakharchenkoPRL09}
    are also shown for comparison.}
  \label{fig:parameterization}
\end{figure}

Here we use a different but more effective way of model parameterization for PFC graphene.
From the previous PFC derivation based on classical DFT \cite{ElderPRB07,HuangPRE10}, to
lowest order the PFC temperature parameter $\epsilon$ can be approximated as
\begin{equation}
  \epsilon = \beta (T-T_m)/T_m,
  \label{eq:epsilon}
\end{equation}
where $\beta$ is the proportional coefficient that depends on the specific material, and
$T_m$ is the melting temperature at which $\epsilon=0$. For graphene the value of $T_m$
has been calculated from atomistic MC or MD simulations, but the result depends on the
interaction potential and the model used, ranging from $4510$ K \cite{LosPRB15} to
$5500$ K \cite{SinghPRB13}. Here we choose the more recent result of $T_m=4510$ K which
was obtained through the combination of atomistic MC simulation using LCBOPII potential
and nucleation theory analysis \cite{LosPRB15}. To estimate $\beta$ for graphene we make
use of a result from atomistic MC \cite{ZakharchenkoPRL09} and MD \cite{doi:10.1063/1.3488620}
simulations of pristine monolayer graphene, both showing a small variation of isothermal
Young's modulus $Y$ within the temperature range of $0 \leq T < 1500$ K followed by
a decrease of $Y$ value with the increase of temperature. A similar behavior of
temperature dependence is reproduced in our PFC simulations of graphene single crystals
(see Fig.~\ref{fig:parameterization}). The simulation setup is similar to that given in
Ref.~\cite{PhysRevE99013302} (other than a perfect single crystal here in the solid region
without double notches). We have tested different system sizes with uniaxial tensile loading
along either the armchair or zigzag direction, all of which indicate a turning point
around $\epsilon=-1.3$ (corresponding to $T \simeq 1500$ K) for the decrease of $Y$.
Thus $\beta = 1.3/(1-1500/T_m) = 1.94784053$, and from Eq.~(\ref{eq:epsilon})
\begin{equation}
  T = (1 + \epsilon/\beta) T_m 
  = T_m + (T_m-1500) \frac{\epsilon}{1.3}, \label{eq:T_conversion}
\end{equation}
which is used for the temperature unit conversion in this work. In the above GB study
the values of $\epsilon$ are chosen from $-0.5$ to $-0.1$, corresponding to temperature
$T$ ranging from $3352$ K to $4278$ K. Given the condition $T \geq 0$, from
Eq.~(\ref{eq:T_conversion}) we can determine the physically meaningful range of
PFC temperature parameter as $\epsilon \geq -\beta = -1.94784053$ for graphene.

The PFC values for Young's modulus and stress can also be converted to real units by
noting the scaling
\begin{equation}
  Y / Y_2^{\rm PFC} = \sigma / \sigma^{\rm PFC} \equiv c_Y,
\end{equation}
where $Y_2^{\rm PFC}$ is the 2D PFC Young's modulus and $\sigma^{\rm PFC}$ is the dimensionless
stress calculated in PFC. Knowing that the lowest temperature used in our single-crystal PFC
simulation is at $\epsilon=-1.9$ with $T \simeq 111$ K, and the corresponding PFC Young's
modulus $Y_2^{\rm PFC}(\epsilon=-1.9) = 1.83126$ (for the armchair direction; see
Fig.~\ref{fig:parameterization}) can be matched to the low-temperature value of $Y=1$ TPa
measured in experiments \cite{Lee385}, we have the conversion factor $c_Y = 1/1.83126$ TPa,
which is applied to all our results of mechanical calculations.

A sample result of our PFC calculation for Young's modulus of pristine graphene is presented
in Fig.~\ref{fig:parameterization}, with system size $2048 \Delta x \times 2048 \Delta y$
(i.e., $54.5$ nm $\times$ $54.5$ nm) and the initial active solid zone of size $1648 \Delta x
\times 1648 \Delta y$ (i.e., $43.9$ nm $\times$ $43.9$ nm). Both the dimensionless PFC values
and the corresponding real-unit converted values are shown in the figure, as well as the data
from atomistic MC calculations in Ref.~\cite{ZakharchenkoPRL09} which used the same LCBOPII
potential as that of Ref.~\cite{LosPRB15} giving $T_m=4510$ K. Note that overall the MC data
points appear to be shifted upward (i.e., of larger value) compared to the PFC ones, which
is due to the use of $Y = 1$ TPa at low $T$ in our unit scaling based on the experimental
result \cite{Lee385}, instead of $1.06$ TPa calculated in Ref.~\cite{ZakharchenkoPRL09}.

The conversion for length scale can be determined by \cite{PhysRevLett.118.255501}
\begin{equation}
  l / l^{\rm PFC} = a_0 / a_0^{\rm PFC} \simeq (2.46~\textrm{\AA}) / (4\pi / \sqrt{3})
  = 0.3391~\textrm{\AA},
\end{equation}
based on the PFC lattice spacing $a_0^{\rm PFC} \simeq 4\pi / \sqrt{3}$ and the graphene lattice
constant $a_0 = 2.46$ {\AA}. Thus in the above PFC simulations of GBs the system size used,
$512\Delta x \times 2048\Delta y$, corresponds to $13.6$ nm $\times$ $54.5$ nm, while its
initial active solid zone is of size $512\Delta x \times 1248\Delta y$, i.e., $13.6$ nm
$\times$ $33.2$ nm. In addition, the time scale of PFC has been identified (see
Ref. \cite{PhysRevE99013302}), i.e.,
\begin{equation}
  t / t^{\rm PFC} = \left ( a_0 / a_0^{\rm PFC} \right )^2 (D^{\rm PFC} / D),
  \label{eq:timescale}
\end{equation}
where $D^{\rm PFC}$ is the dimensionless vacancy diffusion constant in PFC \cite{PhysRevE.70.051605}
and $D$ is the corresponding value of the real material (i.e., graphene). This clearly indicates
the diffusive nature of the characteristic timescales for evolution and dynamics of PFC.

For completeness we summarize in Table \ref{table:GBs} a list of six GBs examined in this
work, including the corresponding type of dislocations at the GB and the grain boundary
energies calculated from the PFC model. Values of these GB energies are adapted from
Ref.~\cite{PhysRevB.94.035414} which also presents the results across the full range of
misorientation angles for various versions of PFC models as well as MD and quantum DFT
calculations.

\begin{table}
  \caption{Summary of the six symmetric tilt grain boundaries studied in this work, including
    the type of dislocations at the boundary and the grain boundary energy which has been
    calculated in Ref.~\cite{PhysRevB.94.035414} using the same PFC model as that of
    Sec.~\ref{sec:model}.}
  \centering
  \begin{tabular}{cccc}
    \hline \hline
    & $\theta$ & Dislocation type & GB energy (eV/nm) \cite{PhysRevB.94.035414}\\
    \hline
    AC1 & $7.34^\circ$ & $(1, 0)$ & 4.2450 \\
    AC2 & $13.17^\circ$ & $(1, 0)$ & 5.6655 \\
    AC3 & $21.79^\circ$ & $(1, 0)$ & 6.6075 \\
    ZZ1 & $8.61^\circ$ & $(1, 0), (0, 1)$ & 5.2678 \\
    ZZ2 & $13.17^\circ$ & $(1, 0), (0, 1)$ & 6.3764 \\
    ZZ3 & $23.49^\circ$ & $(1, 0) + (0, 1) + (1, 0)$, & 7.4849 \\
    & &  $(0, 1) + (1, 0) + (0, 1)$ & \\
    \hline
    \hline
  \end{tabular}
  \label{table:GBs}
\end{table}

\section{Numerical scheme and low-temperature GB results}
\label{sec:numerical}

In this work we use a pseudospectral algorithm to solve the PFC dynamic Eq.~(\ref{1106}).
The equation can be rewritten as $\partial \phi / \partial t = \tilde{L}\phi + f$, where
$\tilde{L}$ represents a linear operator given by $\tilde{L}=\nabla^2[\epsilon+(1+\nabla^2)^2]$,
and the nonlinear terms are denoted as $f=\nabla^2(\tau\phi^2+\phi^3)$. This is a sixth-order
nonlinear partial differential equation. If transforming the equation to Fourier space, with
$\tilde{\phi}_k=\int d\vec{r}\phi(\vec{r})e^{-i\vec k\cdot\vec r}$ the Fourier transform of the
density field $\phi$, the PFC equation can be simplified as
\begin{equation}
  \frac{\partial \tilde{\phi}_k}{\partial t} = \tilde{L}_k \tilde{\phi}_k + \tilde{f}_k,
\end{equation}
where $\tilde{L}_k=-(\epsilon+1)k^2+2k^4-k^6$ and $\tilde{f}_k= -k^2 \int d \vec{r} (\tau\phi^2+\phi^3)
e^{-i\vec k\cdot\vec r}$ represent the Fourier transforms of the linear operator and the nonlinear terms,
respectively, and $\vec k$ is the wave vector. Following the exponential propagation scheme for
the linear term and the predictor-corrector method for the nonlinear terms as described in
Ref.~\cite{doi:10.1063/1.166038}, we have
\begin{eqnarray}\label{1010}
  &\tilde{\phi}_k(t+\Delta t)=&e^{\tilde{L}_k\Delta t} \tilde{\phi}_k(t)
  +\frac{e^{\tilde{L}_k\Delta t}-1}{\tilde{L}_k}\tilde{f}_k(t) \nonumber\\
  && +\left ( e^{\tilde{L}_k\Delta t}-\tilde{L}_k\Delta t-1 \right )
  \frac{\tilde{f}_k(t+\Delta t)-\tilde{f}_k(t)}{{\tilde{L}_k}^2\Delta t},
\end{eqnarray}
when $\tilde{L}_k\ne 0$, and
\begin{equation}\label{1011}
  \tilde{\phi}_k(t+\Delta t)=\tilde{\phi}_k(t)+\tilde{f}_k(t)\Delta t
  +\frac{1}{2}\Delta t \left [\tilde{f}_k(t+\Delta t)-\tilde{f}_k(t) \right ],
\end{equation}
when $\tilde{L}_k=0$. This implicit scheme can be implemented by an iterative procedure of
predictor-corrector steps, although we usually need only one iteration of the corrector step
for numerical convergence and accuracy.

\begin{figure}
  \centering
  \includegraphics[width=0.48\textwidth]{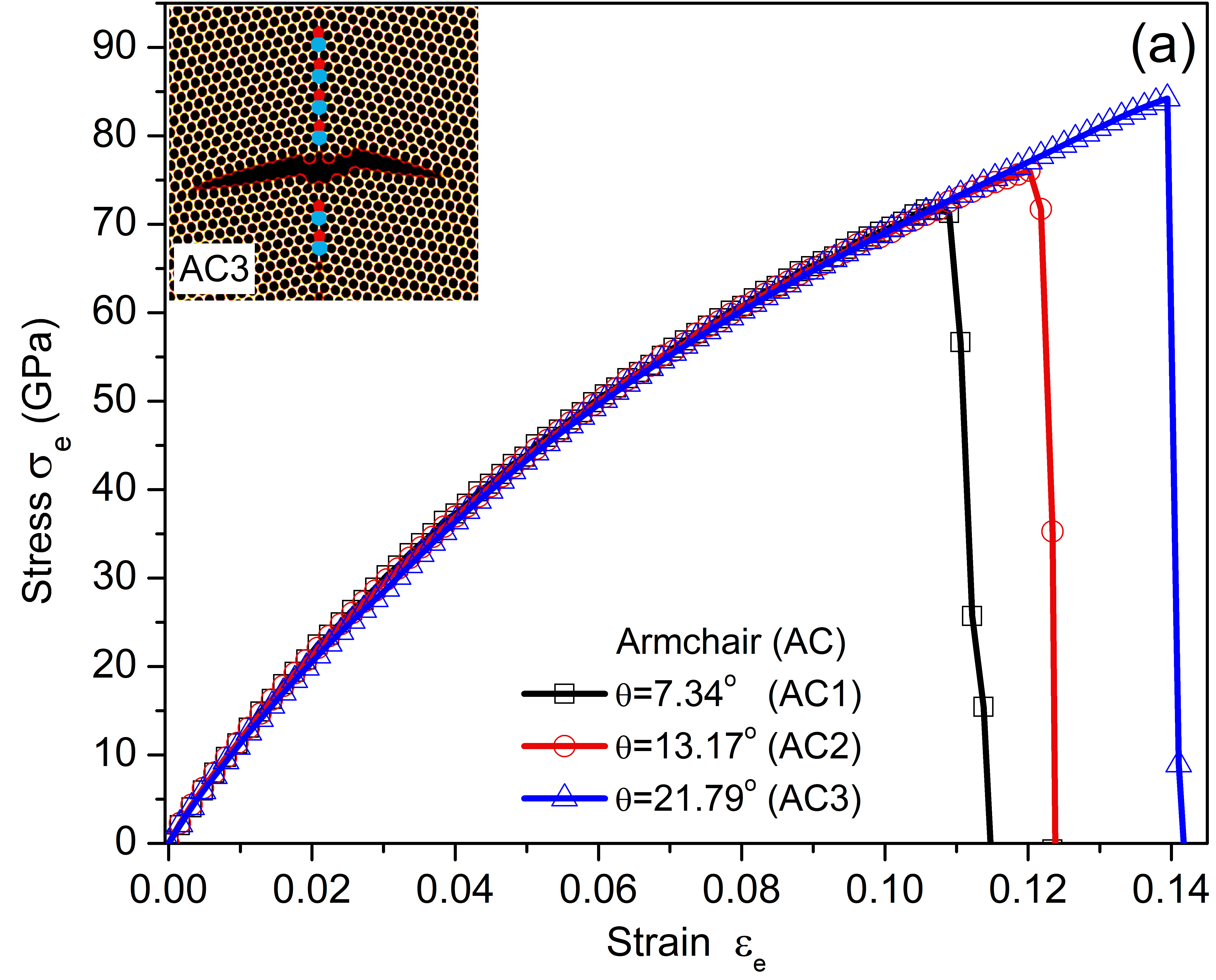}
  \includegraphics[width=0.49\textwidth]{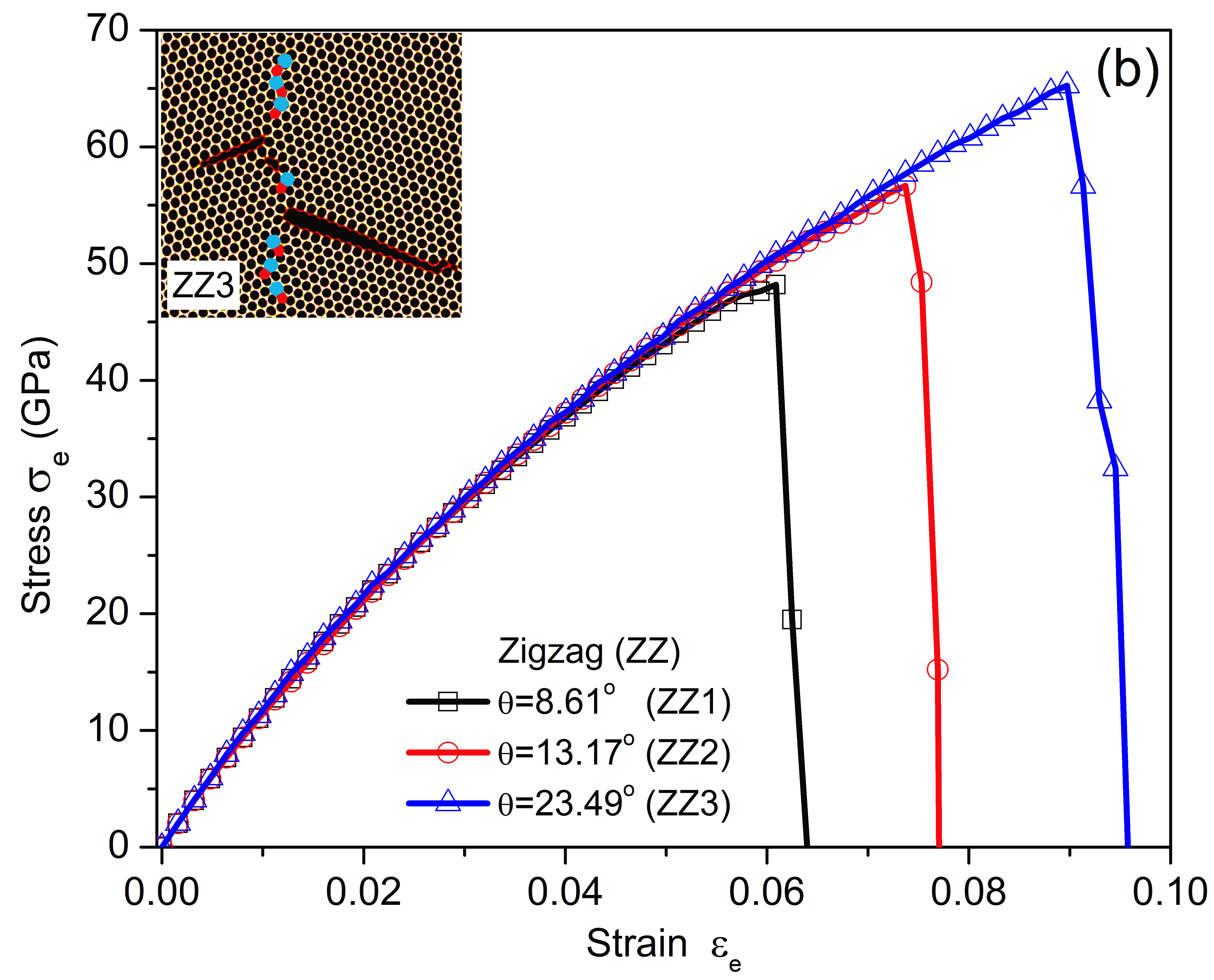}
  \caption{Stress-strain curves for (a) AC and (b) ZZ sheets at $\epsilon=-1.9$ ($T=111$ K)
    under uniaxial tension applied parallel to the GBs. Sample atomic structures at the
    stage of brittle fracture and cracking are also shown in the insets of (a) for AC3
    at $\varepsilon_e=14.26\%$ and (b) for ZZ3 at $\varepsilon_e=9.46\%$.}
  \label{fig:lowT}
\end{figure}

In addition to the high-temperature simulations given above, we have conducted mechanical
calculations for both AC and ZZ GBs at a low temperature of $\epsilon=-1.9$ ($T=111$ K),
to compare with previous atomistic simulations and experimental observation. Other model
parameters and system setup are the same as those in Secs.~\ref{sec:mech_brittle} and
\ref{sec:mech_highT} (except for a smaller time step $\Delta t=0.1$). Some results are
presented in Fig.~\ref{fig:lowT} for tensile loads parallel to the GB, including the
stress-strain curves and some snapshots of brittle fracture and crack propagation into the
grain interior (but not along the GB), which are consistent with those of previous low-temperature
MD simulations \cite{Grantab946} and experiments \cite{KimNanoLett12,RasoolNatCommun13} of
graphene. Note that the notation of armchair vs. zigzag GB used here is opposite to that in
Ref.~\cite{Grantab946}; also the PFC stress values obtained here for ZZ GBs (Fig.~\ref{fig:lowT}(b))
are noticeably lower than the MD results, which can be attributed to the fact that the dislocation
structure and arrangement of ZZ GBs that form naturally from PFC dynamic evolution of density
field are more disperse than those set up manually or by a predetermined way in MD simulations
and are of lower energy (see also the discussion at the end of Sec.~\ref{sec:mech_brittle}).

\bibliographystyle{elsarticle-num} 
\bibliography{mybibfile}

\end{document}